\documentclass[preprint,12pt,review]{elsarticle}

\usepackage{amssymb}
\usepackage{amsthm}
\usepackage{amsmath}
\usepackage{mathrsfs}
\usepackage{graphicx}
\usepackage{epstopdf}
\usepackage{float}
\usepackage{caption}
\usepackage{subcaption}
\usepackage{bm}
\usepackage{amssymb} %for arrows
\usepackage{extarrows} %for arrows
\usepackage{bbm}
\usepackage{mathrsfs}
\usepackage{soul}
\usepackage{hyperref}
\usepackage{cleveref}
\usepackage{siunitx}
\usepackage{tabu} % To increase the vertical space between cells in tables
\usepackage{longtable}
\usepackage{xcolor} % To use different colors in the fonts
\usepackage{comment} % To comment large blocks of text
\usepackage[margin=2cm]{geometry}% by courtesy of Mico
\biboptions{sort&compress} % to show citations as [1-3] instead of [1,2,3]
\usepackage{upgreek} %For non-italic greek letter
\usepackage{soul}
\journal{Computer Methods in Applied Mechanics and Engineering}

\begin{document}

%%%%%%%%%%%%%%%%%%%%%%%%%%
%%%%%%%%%%%%%%%%%%%%%%%%%%
%%%%%%%%%%%%%%%%%%%%%%%%%

%% Abstract document
%%%%%%%%%%%%%%%%%%%%%%%%%%%%%%%%%%%%%%%%%%%%%%%%%%%%
%%%%%%%%%%%%%%%%%%%%%%%%%%%%%%%%%%%%%%%%%%%%%%%%%%%%
%%0.Abstract
%%%%%%%%%%%%%%%%%%%%%%%%%%%%%%%%%%%%%%%%%%%%%%%%%%%%
%%%%%%%%%%%%%%%%%%%%%%%%%%%%%%%%%%%%%%%%%%%%%%%%%%%%
\begin{frontmatter}

% Authors
\title{Electromechanical phase-field fracture modelling of piezoresistive CNT-based composites}

\author[IC]{Leonel Quinteros}
%\ead{l.quinteros-palominos20@imperial.ac.uk}

\author[IC,UGR]{Enrique Garc\'{i}a-Mac\'{i}as}
%\ead{enriquegm@ugr.es}

\author[IC]{Emilio Mart\'{\i}nez-Pa\~neda\corref{cor1}}
\ead{e.martinez-paneda@imperial.ac.uk}
\address[IC]{Department of Civil and Environmental Engineering, Imperial College London, London SW7 2AZ, UK}
\address[UGR]{Department of Structural Mechanics and Hydraulic Engineering, University of Granada, Av. Fuentenueva sn 18002, Granada, Spain}
\cortext[cor1]{Corresponding author.}

\begin{abstract}
We present a novel computational framework to simulate the electromechanical response of self-sensing carbon nanotube (CNT)-based composites experiencing fracture. The computational framework combines electrical-deformation-fracture finite element modelling with a mixed micromechanics formulation. The latter is used to estimate the constitutive properties of CNT-based composites, including the elastic tensor, fracture energy, electrical conductivity, and linear piezoresistive coefficients. These properties are inputted into a coupled electro-structural finite element model, which simulates the evolution of cracks based upon phase-field fracture. The coupled physical problem is solved in a monolithic manner, exploiting the robustness and efficiency of a quasi-Newton algorithm. 2D and 3D boundary value problems are simulated to illustrate the potential of the modelling framework in assessing the influence of defects on the electromechanical response of meso- and macro-scale smart structures. Case studies aim at shedding light into the interplay between fracture and the electromechanical material response and include parametric analyses, validation against experiments and the simulation of complex cracking conditions (multiple defects, crack merging). The presented numerical results showcase the efficiency and robustness of the computational framework, as well as its ability to model a large variety of structural configurations and damage patterns. The deformation-electrical-fracture finite element code developed is made freely available to download.\\
\end{abstract}

\begin{keyword}
Carbon nanotubes (CNTs) \sep  Finite element analysis \sep Phase-field \sep Piezoresistivity  \sep  Smart materials \sep  Fracture 
\end{keyword}

%\begin{extrainfo}
%\textbf{Highlights} % Less than 85 characters, blank spaces included
%\begin{itemize}
%\item A novel smart beam concept is proposed for buckling detection applications
%\item A micromechanics-based FEM is used to model macroscopic CNT-based sensors
%\item Electromechanical constitutive properties are benchmarked against experimental data
%\item Buckling is shown to be traceable from the electrical output of the sensors
%\item Study on the electrodes layout is presented, including their length and position
%\end{itemize}
%\end{extrainfo}

\end{frontmatter}

%\linenumbers

%%%%%%%%%%%%%%%%%%%%%%%%%%%%%%%%%%%%%%%%%%%%%%%%%%%%
%%%%%%%%%%%%%%%%%%%%%%%%%%%%%%%%%%%%%%%%%%%%%%%%%%%%
%%1.Introduction
%%%%%%%%%%%%%%%%%%%%%%%%%%%%%%%%%%%%%%%%%%%%%%%%%%%%
%%%%%%%%%%%%%%%%%%%%%%%%%%%%%%%%%%%%%%%%%%%%%%%%%%%%
\newcommand\boldred[1]{\textcolor{red}{\textbf{#1}}}
\section{Introduction}
\label{section 1:Introduction}

Recent advances in the development of nano-modified multifunctional materials such as self-sensing CNT-based composites have opened vast new possibilities in the realm of Structural Health Monitoring (SHM). These include their use in laminated composites with superior stiffness/weight ratios for aeronautical structures~\cite{Zhang2015,VERTUCCIO2016192}, self-diagnostic concretes~\cite{DALESSANDRO2016200,GARCIAMACIAS201745,s18030831}, and smart clothing applications~\cite{Yamada2011}, just to mention a few. Among the multifunctional properties of CNT-based composites, their piezoresistive properties have garnered particular interest among the scientific community due to their potential for the development of next-generation self-diagnostic materials. When doping small dosages of CNTs in polymer or cementitious materials, the resulting composite exhibits strain self-sensing properties through a piezoresistive effect~\cite{CAO20171,Hu2010,Ayesha2016,Vadlamani2010}. This enables the development of smart load-bearing sensors capable of monitoring its own strain condition through electrical resistivity measurements~\cite{birgin2020}.

An essential step to model piezoresistive CNT-based composites consists in estimating their electromechanical constitutive properties. To this aim, several approaches have been proposed in the literature, two of the most successful methods being molecular dynamics (MD)~\cite{FRANKLAND20031655,GRIEBEL20041773} and first principles-based approaches~\cite{Natsuki2004,GARCIAMACIAS2019114}. However, these atomic level calculations are limited in the time and space scales that can be addressed. An attractive and computationally efficient solution relies on mean-field homogenisation (MFH), which can simulate large and complex composite microstructures in an analytical or semi-analytical fashion. In addition, multiple micromechanical features can be incorporated in the simulation, including the geometrical properties of CNTs, filler waviness, agglomeration and orientation distribution~\cite{GARCIAMACIAS201849,GARCIAMACIAS2017208}. In this light, Hori and Nemat-Nasser~\cite{Hori1993} calculated the elastic moduli of composites filled with nano-inclusions using the Mori-Tanaka homogenisation theory, considering the existence of an interphase coating between the matrix and the inclusions. Xu \textit{et al.}~\cite{XU2017162} estimated the volume fraction of both soft and hard interphases for ellipsoidal inclusions. MFH approaches have also been applied to estimate the electrical conductivity and the strain self-sensing properties of CNT-based composites. Experiments have shown that CNT-based composites exhibit two main conduction mechanisms as a result of their percolation-like nature~\cite{WEN2007263,CHIARELLO2005463}: conductive networking and electron hopping (or quantum tunnelling)~\cite{GOVOROV2018174,WENTZEL201763}. Below a critical volume fraction known as the percolation threshold, the conductive fillers are distant from each other and electrons can only be transferred by trespassing the potential barrier of the matrix through a quantum tunnelling mechanism. As the concentration of fillers approaches the percolation threshold, CNTs get in contact with each other creating a continuous conductive path. The latter, also referred to as the conductive networking mechanism, results in sudden increases in the electrical conductivity of the composite, which may be several orders of magnitude higher than that of the pristine matrix phase. Feng and Jiang~\cite{FENG2013143} built upon this physical understanding to estimate the electrical conductivity of CNT-polymer composites through a micromechanics approach. A similar strategy was followed by Garc\'{i}a-Mac\'{i}as \textit{et al.}~\cite{GARCIAMACIAS2017451,GARCIAMACIAS2017195} to estimate the electrical conductivity and linear piezoresistivity coefficients of cement-based composites doped with CNTs. In regard to the fracture behaviour of CNT-based composites, experiments have revealed a notable toughening effect due to CNT bridging mechanisms~\cite{MIRJALILI20101537}. Micromechanical models have thus been enriched to account for CNT bridging mechanisms such as CNT pull-out and rupture \cite{FU19961179,menna2016effect}.

While considerable efforts have been exerted to estimate the constitutive properties of CNT-based composites, the number of works addressing the role of defects on the electromechanical response of meso- and macro-scale CNT-based composites is considerably scarce. Negi~\textit{et al.}~\cite{Negi2019} used the extended finite element method (X-FEM) to predict the influence of crack-like defects upon the mechanical response of composite plates doped with CNTs. Downey \textit{et al.}~\cite{DOWNEY2017924} proposed a resistor network model to approximate the electric field in CNT-cement composites and conduct damage detection, localization and quantification. Rodr\'{i}guez-Tembleque~\textit{et al.}~\cite{RODRIGUEZTEMBLEQUE2020102470} used X-FEM to investigate the role of cracks on the electrical output of CNT-based composite sensors. Exploiting the one-way coupling of piezoresistive materials, their approach employed a sequential two-step procedure~\cite{RODRIGUEZTEMBLEQUE2022115137}: the mechanical problem is solved first, and the resulting strain field is used to update the local electrical conductivity of the material and subsequently obtain a solution for the electrical field. Despite these encouraging results, X-FEM is known to suffer some computational limitations; these include the need for a-priori definition of the location and orientation of the crack, challenges in handling 3D boundary value problems, and difficulties when dealing with multiple interacting cracks, to mention a few. Alternatively, the so called phase-field fracture method has been proposed as a powerful technique to simulate complex cracking phenomena in arbitrary geometries and dimensions~\cite{Borden2012,Borden2014,Borden2016}. Phase-field fracture modelling has attained remarkable popularity in recent years due to its robustness, ease of implementation, high flexibility in simulating complex problems (crack branching, merging, complex trajectories), and straightforward integration in coupled physical simulations \cite{Wu2020}. Evidence of this is found in the multiple recent applications of the phase-field to a wide variety of materials and fracture phenomena, including hydrogen embrittlement~\cite{MARTINEZPANEDA2018742,Wu2020b,Dinachandra2022}, shape memory alloys~\cite{Simoes,SIMOES2021113504}, composite materials~\cite{TAN2022115242,Quintanas-Corominas2020a}, iceberg calving \cite{Sun2021,Clayton2022}, and Li-ion batteries~\cite{Zhang2016,Boyce2022,Ai2022}. In the realm of CNT-based composites, a combined micromechanics and phase-field fracture framework was very recently proposed by Quinteros~\textit{et al.}~\cite{QUINTEROS2022109788}. The work incorporated for the first time the main bridging mechanisms and showcased the ability of the proposed framework to capture the sensitivity of the fracture resistance to microstructural aspects such as filler aspect ratio, orientation distribution and agglomeration. However, this work was limited to mechanical phenomena and the fracture-electromechanical interplay is yet to be explored in CNT-based composites. In the context of electromechanical CNT-based composites behaviour, the piezoresistivity effect, and its interplay with fracture, is of particular importance. This is yet to be explored in the context of phase-field fracture modelling as the electromechanical phase-field fracture literature is limited to piezoelectric and ferroelectric materials~\cite{Wilson2013,MIEHE20101716,ABDOLLAHI20122100,WU2021114125}, whereby the influence on fracture of the interplay between electric fields and deformation is assessed. In the context of piezoresistive materials, the interest is on the influence of the mechanical load on the material resistivity and on the degradation of electrical conductivity that results from material damage.

In this work, we present the first computational framework for modelling deformation-electrical-fracture phenomena in CNT-based composites. The framework combines MFH, electromechanical piezoresistivity modelling, and a phase-field description of fracture that accounts for CNT toughening effects. The MFH formulation, used to estimate relevant electrical, mechanical and fracture properties, is presented in Section \ref{section 2: Modelling}. The electromechanical phase-field framework is subsequently described in Section \ref{Section:3}, including details of the numerical implementation and the interaction between electrical permeability/conductivity and phase-field damage. The results obtained are shown in Section \ref{Section:4-Results}. Five case studies are investigated, which include an experimental validation and the simulation of various electrical-cracking phenomena in 2D/3D geometries containing multiple defects. Finally, the manuscript ends with concluding remarks in Section \ref{Section:5-Conclusion}. 

%%%%%%%%%%%%%%%%%%%%%%%%%%%%%%%%%%%%%%%%%%%%%%%%%%%%
%%%%%%%%%%%%%%%%%%%%%%%%%%%%%%%%%%%%%%%%%%%%%%%%%%%%
%%2.Mechanical_prop
%%%%%%%%%%%%%%%%%%%%%%%%%%%%%%%%%%%%%%%%%%%%%%%%%%%%
%%%%%%%%%%%%%%%%%%%%%%%%%%%%%%%%%%%%%%%%%%%%%%%%%%%%
\section{Mean-field electromechanical homogenisation of CNT-based composites}
\label{section 2: Modelling}

This section briefly overviews the micromechanics modelling approach used to estimate the constitutive properties of CNT-based composites. Specifically, the modelling of the mechanical and electrical properties are independently presented in Sections \ref{Section 2.1:Mechanical} and \ref{Section 2.3: Electrical properties}, respectively. For the sake of simplicity, CNTs are assumed to be straight, well-dispersed, and randomly oriented. Nevertheless, waviness and agglomeration effects can be readily incorporated in the MFH as shown elsewhere (see e.g.~\cite{QUINTEROS2022109788,GARCIAMACIAS2017451}).

\subsection{Mechanical properties of CNT-based composites}
\label{Section 2.1:Mechanical}

\subsubsection{Elastic tensor}
\label{Section 2.1.1}

\begin{figure}[H]
\centering
\includegraphics[scale=1.15]{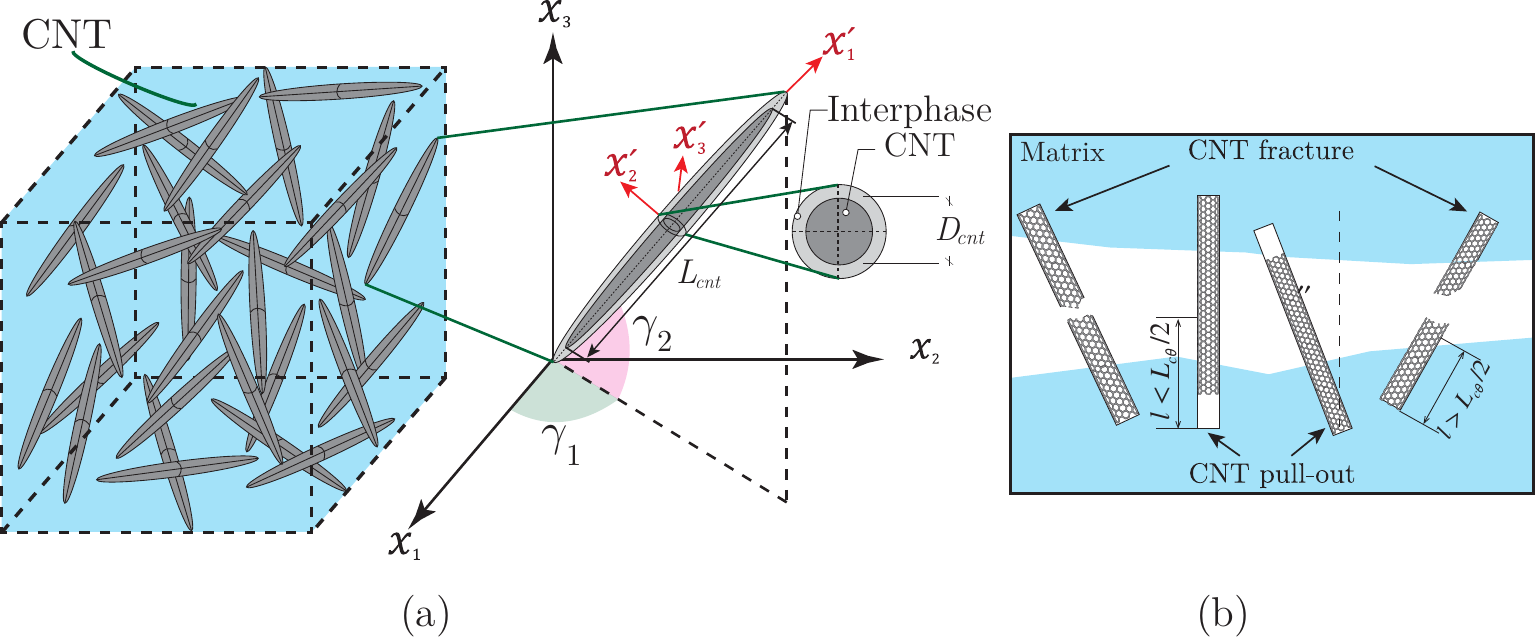}
\caption{CNT-based composites: (a) Sketch of an RVE of a homogeneous matrix material loaded with randomly oriented CNTs, and (b) fracture and toughening mechanisms relevant to CNT-based composites.}
\label{Fig:Fig_1}
\end{figure}

Let us consider a representative volume element (RVE) of a matrix phase doped with CNT as sketched in Fig.~\ref{Fig:Fig_1}a. The RVE is assumed to satisfy three main assumptions: (i) it contains a sufficient amount of CNTs so that the overall properties of the composite are statistically represented; (ii) the CNT length $L_{cnt}$ and diameter $D_{cnt}$ are constant; and (iii) the fillers are randomly oriented. The orientation of each CNT is described with a local coordinate system K$'\equiv\left\{0;x'_1x'_2x'_3\right\}$ defined by two Euler angles $\gamma_1$ and $\gamma_2$. CNTs are taken to be analogous to homogeneous inclusions surrounded by finite elastic coatings with thickness $t$, so as to simulate the matrix/filler load-transfer properties. Therefore, the composite material is defined as a three-phase medium, including the matrix, fillers and interphases, with elastic tensors $\bm{C}_m$, $\bm{C}_p$ and $\bm{C}_i$, respectively. Subscripts $p$, $i$, and $m$ relate the corresponding magnitudes to the filler, interphase and matrix phases, respectively. 
Following the notation of Hori and Nemat-Nasser~\cite{Hori1993}, every CNT and its surrounding interphase is defined as a double inclusion. In this regard, the effective constitutive tensor of the composite $\bm{C}$ can be written as~\cite{XU2017162,GARCIAMACIAS201849}:
\begin{equation}\label{overall}
\bm{C}=\left(f_m\bm{C}_m+f_i\left\langle \bm{C}_i:\bm{A}_i \right\rangle+f_p\left\langle \bm{C}_p:\bm{A}_p  \right\rangle\right):\left(f_m \, \mathbf{I}+f_i\left\langle \bm{A}_i \right\rangle+f_p\left\langle \bm{A}_p \right\rangle\right)^{-1},
\end{equation}

\noindent where $f_p$, $f_i$, and $f_m$ denote the volume fraction of the fillers, the interphases, and the matrix, respectively; the colon operator denotes the tensorial inner product between two tensors, $\left(\bm{A}:\bm{B}\right)_{ijmn}\equiv \bm{A}_{ijkl}\bm{B}_{klmn}$; and tensors $\bm{A}_i$ and $\mathbf{A}_p$ refer to the concentration tensors for interphases and inclusions, respectively. The concentration tensors can be expressed by the corresponding dilute concentration tensors, $\bm{A}_i^{dil}$ and $\bm{A}_p^{dil}$, as:
\begin{equation}
\bm{A}_\chi=\bm{A}_\chi^{dil}:\left(f_m\mathbf{I}+f_i\bm{A}_i^{dil}+f_p\bm{A}_p^{dil}\right)^{-1},\quad \chi=p,i
\end{equation}
\begin{equation}
\bm{A}_\chi^{dil}=\mathbf{I}+\bm{S}:\bm{T}_\chi, \quad \chi=p,i
\end{equation}

\noindent where,
\begin{equation}
\bm{T}_\chi=-\left(\bm{S}+\bm{M}_\chi\right)^{-1}, \quad \chi=p,i
\end{equation}
\begin{equation}
\bm{M}_\chi=\left(\bm{C}_\chi-\bm{C}_m\right)^{-1}:\bm{C}_m, \quad \chi=p,i
\end{equation}

Angle bracket operators $\left\langle \cdot \right\rangle$ in Eq.~(\ref{overall}) represent orientational average, which can be defined for an arbitrary field $\bm{F}$ as:
\begin{equation}
\left\langle \bm{F} \right\rangle=\int_0^{2\pi}\int_0^{\pi/2}\bm{F}(\gamma_1,\gamma_2)\Omega(\gamma_1,\gamma_2)\sin(\gamma_2)\textrm{d}\gamma_2\textrm{d}\gamma_1,
\end{equation}

\noindent where $\Omega(\gamma_1,\gamma_2)$ stands for the orientation distribution function (ODF) of the fillers. In general, CNTs are randomly oriented when dispersed into polymer or cement matrices and the ODF takes the shape of an uniform distribution with a constant value within the whole Euler space, that is $\Omega(\gamma_1,\gamma_2)=1/2\pi$.

Interfacial effects between the CNTs and the matrix must be accounted for, as models neglecting these have shown to overestimate the elastic properties of the composites~\cite{Rafiee2014}. In particular, interfaces are found to constitute weak zones with limited load-transfer properties determined by van der Waals interaction forces. In the realm of CNT-based composites, interfaces can be simulated through compliant penetrable interphases with low stiffness. The formula of the volume fraction of finite soft interphase $f_{i}$ around ellipsoidal particles was derived by Xu \textit{et al}.~\cite{Xu2016} as:
\begin{equation}\label{softvi}
f_i= (1-f_p)\left(1-\exp \left\{-\frac{6f_p}{1-f_p}\left[\frac{\eta}{n(\kappa)}+\left(2+\frac{3f_p}{n^2(\kappa)(1-f_p)}\right)\eta^2 \right.\right.\right. + \left.\left.\left. \frac{4}{3}\left(1+\frac{3f_p}{n(\kappa)(1-f_p)}\right)\eta^3\right]\right\}\right),
\end{equation}

\noindent with $\eta$ being the ratio of the interfacial thickness $t$ to the equivalent diameter $D_{eq}$ (i.e.~$\eta=t/D_{eq}$). The equivalent diameter denotes the diameter of an equivalent sphere with the same volume as the particles~\cite{Beddow2018} and can be determined for CNTs with aspect ratio $\kappa=L_{cnt}/D_{cnt} > 1$ as $D_{eq}=D_{cnt}\kappa^{1/3}$. The term $n(\kappa)$, representing the sphericity of the CNTs, denotes the ratio of the surface area between the equivalent sphere and that of the particles, which is:
\begin{equation}\label{hardvi2}
n(\kappa)=\frac{2\kappa^{2/3}\tan \left(\textrm{arcos} (1/\kappa)\right)}{\tan\left( \textrm{arcos} (1/\kappa)\right)+\kappa^{2}\textrm{arcos} (1/\kappa)}.
\end{equation}

\subsubsection{Fracture energy: Pull-out and rupture}
\label{Sec211}
Nanotubes and nanofibers are responsible for several toughening mechanisms, including debonding, pull-out, and rupture \cite{SUN20092392}. For CNT composites in particular, pull-out and rupture are believed to be the main contributions to fracture resistance \cite{WICHMANN2008329}. Both mechanisms are sketched in Fig.~\ref{Fig:Fig_1}b. Accordingly, the material toughness or critical energy release rate $G_{c}$ can be expressed as the contribution of the matrix material and the CNT bridging mechanisms as~\cite{menna2016effect}:
\begin{equation}
    G_{c}=G_{0}+G_{br}=G_{0}+G_{Po}+G_{Fr}
\end{equation}

\noindent where $G_{0}$ is the matrix fracture energy, and $G_{br}$ is a term encapsulating the two main toughening mechanisms: the rupture and pull-out of CNTs, denoted by $G_{Fr}$ and $G_{Po}$, respectively. CNT pull-out occurs due to the interfacial friction between the CNT and matrix, which develops across the high specific surface area of the nanotubes, and its dominance relative to the CNT rupture mechanism is dependent on the nanotube characteristics and the interfacial bond strength. The pull-out mechanism occurs if the embedment length $l$ of a CNT oriented at a certain angle $\theta$ is equal or lower than a critical length $L_{c\theta}/2$, otherwise CNT rupture will take place~\cite{FU19961179}. The critical length $L_{c\theta}$ is given by:
\begin{equation}
    L_{c\theta}=\frac{\sigma_{ult \theta}D_{cnt}}{2 \tau_{int} \exp \left( \mu \theta \right)},
    \label{Eq:Ltheta}
\end{equation}

\noindent where $\tau_{int}$ is the interfacial shear stress, which can be obtained using atomic force microscopy~\cite{Barber2003}, $\mu$ is the snubbing friction coefficient for misaligned CNTs~\cite{li1991micromechanical}, and $\sigma_{ult\theta}$ is the fracture stress of an oblique CNT, which is given by:
\begin{equation}
  \label{Eq:sigma_ult}
    \sigma_{ult \theta}=\sigma_{ult}(1-A \text{tan}\theta),
\end{equation}

\noindent with $\sigma_{ult}$ being the ultimate tensile strength and $A$ a constant determining the inclined fibre strength. Then, the work done by the pull-out and fracture  of CNTs can be written as a piecewise function of the embedment length $l$ as:
\begin{equation}
\label{Eq:W_en}
W(l,\theta) = \begin{cases}l^{2}\tau_{int}\pi D_{cnt} \exp(\mu \theta)/2 &\mbox{if }l< L_{c\theta}/2 \\
 \pi D_{cnt}^{2} \sigma_{ult}^{2} L_{cnt}/\left(8E_{cnt}\right) & \mbox{if } l\geq L_{c\theta}/2 \end{cases},
\end{equation}

\noindent with $E_{cnt}$ being the Young's modulus of the CNTs. Finally, the fracture energy considering straight CNTs can be obtained as~\cite{fu1997fibre}:
\begin{equation}
    G_{br}= \frac{2f_{p}}{A_{cnt}L_{cnt}}\int_{\theta=0}^{\pi/2} \int_{l=0}^{L_{cnt}/2}  W(l,\theta)g(\theta)\cos(\theta) \, \rm{d}l \rm{d}\theta
    \label{Eq:G_c},
\end{equation}

\noindent where $g(\theta)$ represents the orientation distribution. Despite the orientation of CNTs being eminently three-dimensional, several studies showed that just an angle $\theta$ suffices to describe the orientation between the loading direction and the fibre axis~\cite{FU19961179,jain1992effect}, as illustrated in Fig.~\ref{Fig:Fig_1}b. Then, $g(\theta)$ is defined as~\cite{Xia1995}:
\begin{equation}
    g(\theta)=\frac{[\sin(\theta)]^{2p-1}[\cos(\theta)]^{2q-1}}{\int_{\theta_{min}}^{\theta_{max}} ([\sin (\theta)]^{2p-1}[\cos(\theta)]^{2q-1})\text{d}\theta},
    \label{Eq:orientation_pdf}
\end{equation}

\noindent where $\theta_{min}\leq \theta \leq \theta_{max}$ and $p\geq 1/2$, $q\geq 1/2$ are parameters that determine the shape of the distribution. 

\subsection{Electrical properties of CNT-based composites}
\label{Section 2.3: Electrical properties}

\subsubsection{Electrical conductivity}
\label{Section 2.3.1: Conductivity}

\begin{figure}[H]
\centering
\includegraphics[scale=0.85]{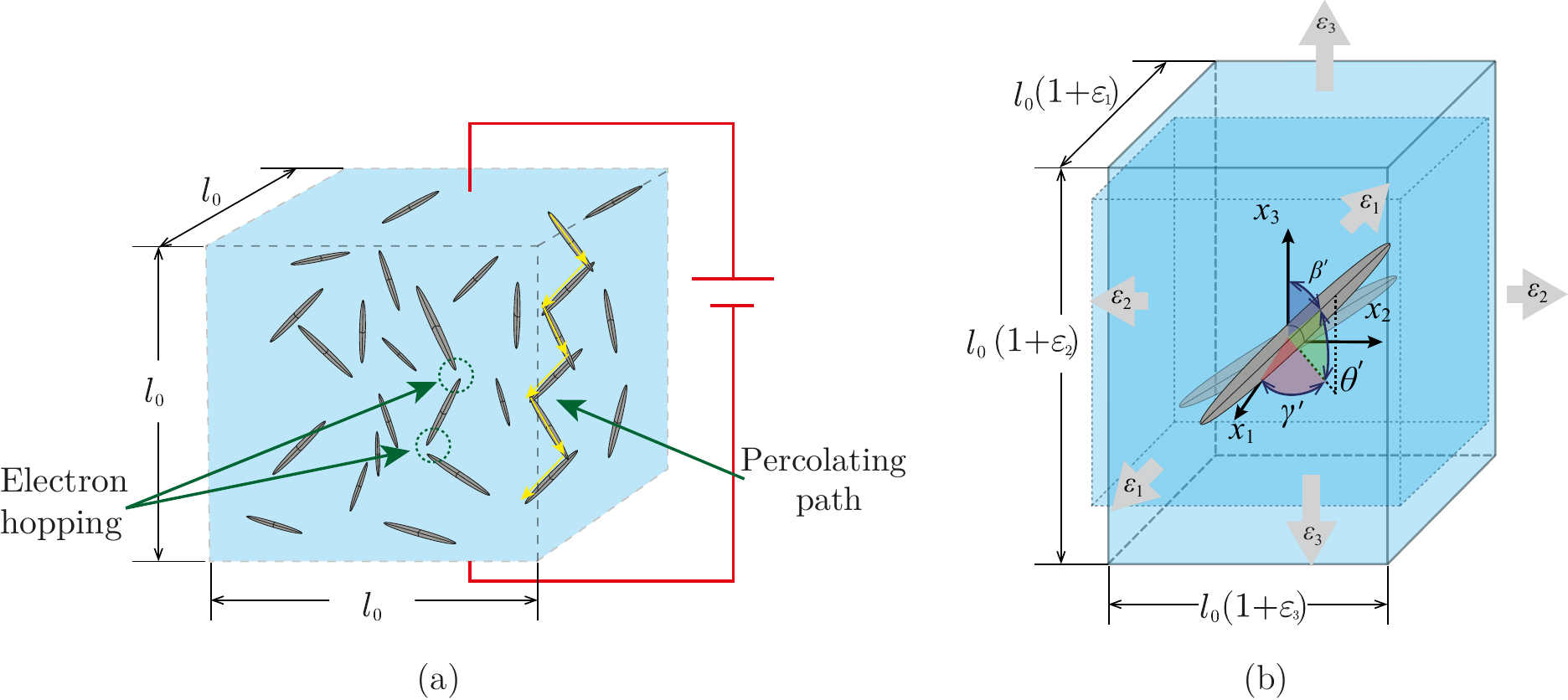}
\caption{Schematic illustration of: (a) the electron hopping and conductive networking mechanisms governing the overall electrical conductivity of CNT-based composites, and (b) the strain-induced filler reorientation effect, in a $l_{0} \times l_{0} \times l_{0}$ cubic deformable cell.}
\label{Fig:Fig_2}
\end{figure}

The modelling of the electrical conductivity of CNT-based composites follows a similar micromechanical procedure to the one previously presented in Section~\ref{Section 2.1:Mechanical}. Note that $f_{p}(\bm{\varepsilon})$ and $f_{c}(\bm{\varepsilon})$ are both functions of the mechanical strain, whose dependency will be explained later. Percolation theory indicates that the electrical conductivity mechanism depends on the filler volume fraction in a non-linear way. Specifically, if the CNT filler content $f_{p}(\bm{\varepsilon})$ is below the percolation threshold $f_{c}(\bm{\varepsilon})$, then the fibres are too distant from each other and electrons can only be transferred through the matrix by a quantum tunnelling effect. However, for filler contents above the percolation threshold, CNTs tend to contact each other forming conductive networks as shown in Fig.~\ref{Fig:Fig_2}a. The fraction of percolated CNTs  can $\xi(\bm{\varepsilon})$ be approximated as~\cite{Deng2008}: 
\begin{equation}
\xi(\bm{\varepsilon})= \begin{cases}0 & 0 \leq f_{p}(\bm{\varepsilon})<f_{c}(\bm{\varepsilon}) \\ \frac{f_{p}(\bm{\varepsilon})^{1 / 3}-f_{c}^{1 / 3}(\bm{\varepsilon})}{1-f_{c}^{1 / 3}(\bm{\varepsilon})} & f_{c}(\bm{\varepsilon}) \leq f_{p}(\bm{\varepsilon}) \leq 1\end{cases}
\label{distinterface}
\end{equation}

In this light, the modelling of the electrical conductivity of CNT-based composites should account for fillers contributing through electron hopping (non-percolating) and conductive networking (percolating mechanisms). This can be formalised through MFH as~\cite{GARCIAMACIAS2017195,GARCIAMACIAS2017451,BURONI2021923},
\begin{equation}
    \bm{\sigma}_{eff}(\bm{\varepsilon}) = \bm{\sigma}_{m} + \bm{\sigma}_{NP}(\bm{\varepsilon}) +  \bm{\sigma}_{P}(\bm{\varepsilon}),
    \label{Eq:Electrical_cond}
\end{equation}
\begin{equation}
    \bm{\sigma}_{P}(\bm{\varepsilon}) = \xi(\bm{\varepsilon}) \left\langle f_{p}(\bm{\varepsilon})(\bm{\sigma}_{CNT}^{P}(\bm{\varepsilon})-\bm{\sigma}_{m}) \mathbf{A}_{P}\right\rangle,
\end{equation}
\begin{equation}
    \bm{\sigma}_{NP}(\bm{\varepsilon}) = (1-\xi(\bm{\varepsilon}))\left\langle f_{p}(\bm{\varepsilon}) (\bm{\sigma}_{CNT}^{NP}(\bm{\varepsilon})-\bm{\sigma}_{m}) \mathbf{A}_{NP}\right\rangle,
\end{equation}

\noindent where the terms $\bm{\sigma}_{m}$, $\bm{\sigma}_{NP}$ and $\bm{\sigma}_{P}$ correspond to the conductivity tensor of the matrix, the non-percolating (NP) contribution and the percolating (P) contribution, respectively. The tensors $\bm{\sigma}_{NP}$ and $\bm{\sigma}_{P}$ are calculated using the transversely isotropic electrical conductivity tensor of an equivalent solid cylinder accounting for the electrical conductivity of CNTs and the surrounding volume of matrix material, where electron hopping may develop (refer to~\ref{Apendix_A} for further details). The quantities in $\bm{\sigma}_{NP}$ are computed assuming the aspect ratio of a prolate ellipsoid with $a_{2}=a_{3}=D_{cnt}/2$ and $a_{1}=L$, while the terms in $\bm{\sigma}_{P}$ consider $a_{2}=a_{3}=r_{c}$ and $a_{1} \rightarrow$ $\infty$. The concentration tensor $\mathbf{A}_{j}$, with $j$ being $NP$ or $P$ is estimated as:
\begin{equation}
    \mathbf{A}_{j}=\mathbf{A}^{dil}\left( (1-f_{p}(\bm{\varepsilon}))\mathbf{I} + f\mathbf{A}^{dil}\right)^{-1}, \quad \mathbf{A}^{dil}=(\mathbf{I}+\mathbf{S} \bm{\sigma}_{m}^{-1}(\bm{\sigma}_{f}-\bm{\sigma}_{m}))^{-1},
\end{equation}

\noindent where $\mathbf{I}$ is the $3 \times 3$ identity matrix and $\mathbf{S}=\text{diag}(S_{11},S_{11},S_{33})$ is the shape-dependent Eshelby's tensor. The components of $\mathbf{S}$ are given by~\cite{Eshelby1957,Eshelby1959}:
\begin{equation}
    S_{11}=\frac{s}{2(s^{2}-1)^{3/2}}\left[s(s^{2}-1)^{1/2} - \cosh^{-1}(s) \right], \quad S_{33}=1-2S_{11},
    \end{equation}

\noindent with $s$ being the filler aspect ratio.

An important aspect of Eq.~\eqref{Eq:Electrical_cond} is the conductivity dependence on the strain tensor $\bm{\varepsilon}$, which induces a piezoresistive effect into the composite~\cite{BURONI2021923}. Three main mechanisms are consistently identified in the literature driving such an effect, namely: (i) volume expansion, (ii) filler reorientation, and (iii) variation of the percolation threshold. The first mechanism considers that strains alter the volume fraction of the fillers $f_{p}(\bm{\varepsilon})$, which in turn modifies the fraction of percolated fillers $\xi$. Following the formulation by Garc\'{i}a-Mac\'{i}as \textit{et al.}~\cite{Garcia-Macias2018a}, the dependency between $f_{p}$, the unstrained fibre volume fraction $f_{p}(0)$, and a general strain state $\bm{\varepsilon}$ reads:
\begin{equation}
    f_{p}(\bm{\varepsilon})=\frac{f_{p}(0)}{\bar{\varepsilon}_{1}\bar{\varepsilon}_{2}\bar{\varepsilon}_{3}} = \frac{f_{p}(0)}{\text{tr}(\bm{\varepsilon})+\det(\bm{\varepsilon})\text{tr}(\bm{\varepsilon}^{-1})+\det(\bm{\varepsilon})+1}.
\end{equation}

\noindent with $\bar{\varepsilon}_{1}=\varepsilon_{1}+1$, $\bar{\varepsilon}_{2}=\varepsilon_{2}+1$ and $\bar{\varepsilon}_{3}=\varepsilon_{3}+1$. Here, $\varepsilon_{1}$, $\varepsilon_{2}$, $\varepsilon_{3}$ are the three principal strains, and $\det(\cdot)$ and $\text{tr}(\cdot)$ denote the determinant and trace operators, respectively. The filler reorientation induced by mechanical strain is sketched in Fig.~\ref{Fig:Fig_2_1}b. Note that mechanical strains tend to decrease the randomness in the distribution of the fillers orientation. Under the assumption of rigid rotations of inextensible fibres, Garc\'{i}a-Mac\'{i}as and co-authors~\cite{Garcia-Macias2018a} described the relationship between an arbitrary strain state $\bm{\varepsilon}$ and the ODF to be used in the orientational averages in Eq.~\eqref{Eq:Electrical_cond} as:
\begin{equation}  w(\bar{\varepsilon}_{1},\bar{\varepsilon}_{2},\bar{\varepsilon}_{3}|\gamma_{1},\gamma_{2})=\frac{\bar{\varepsilon}_{1}^{2}\bar{\varepsilon}_{2}^{2}\bar{\varepsilon}_{3}^{2}}{\left[\bar{\varepsilon}_{1}^{2}\bar{\varepsilon}_{2}^{2}\cos^{2}(\gamma_{2})+\bar{\varepsilon}_{3}^{2} \left( \bar{\varepsilon}_{1}^{2}\sin^{2}(\gamma_{1})+\bar{\varepsilon}_{2}^{2}\cos^{2}(\gamma_{1}) \right) \sin ^{2}(\gamma_{2})\right]^{3/2}}
    \label{Eq:w_distribution_eq}
\end{equation}

Finally, the reorientation of fillers induces a change in the percolation threshold, which can be formulated following the percolation theory set out by Komori and Makishima~\cite{KomoriMakishima1977} as:
\begin{equation}
 f_{c} =\frac{\pi}{5.77 s I},
\end{equation}
\noindent with
\begin{equation}
    I=\int_{0}^{\pi} \int_{0}^{\pi} J(\gamma_{1},\gamma_{2}) \hat{w}(\gamma_{1},\gamma_{2}) \sin (\gamma_{2}) \text{d}\gamma_{1} \text{d}\gamma_{2},  
\end{equation}

\begin{equation}
    J(\gamma_{1},\gamma_{2})=\int_{0}^{\pi} \int_{0}^{\pi}\sin \tau (\gamma_{1},\gamma_{1} ' , \gamma_{2},\gamma_{2} ' ) \hat{w}(\gamma_{1}',\gamma_{2}')\sin(\gamma_{2}') \text{d}\gamma_{1} ' \text{d}\gamma_{2} ' ,
\end{equation}

\noindent and
\begin{equation}
    \sin \tau=\left[1-\left(\cos(\gamma_{2})\cos(\gamma_{2}')+\cos(\gamma_{1}-\gamma_{1}')\sin(\gamma_{2})\sin(\gamma_{2}')\right)^{2} \right]^{1/2},
\end{equation}

\noindent where $\hat{w}$ is the normalised ODF from Eq.~\eqref{Eq:w_distribution_eq}.

\subsubsection{Piezoresistivity coefficients}
\label{Section 2.3.2: Piezoresistive}

Under the assumption of small strains, it can be stated that the strain and the electrical resistivity are related by a linear isotropic tensor $\bm{\Pi}$ referred to as the piezoresistivity tensor, such that:
\begin{equation}\label{pimatrix}
\left[ \hspace{-0.6em} \begin{array}{c} \Delta \rho_{11}/\rho_0 \\ \Delta \rho_{22}/\rho_0 \\ \Delta \rho_{33}/\rho_0 \\ \Delta \rho_{23}/\rho_0 \\ \Delta \rho_{13}/\rho_0 \\ \Delta \rho_{12}/\rho_0 \end{array} \hspace{-0.6em} \right]=
\begin{bmatrix} 
\lambda_{11} &  \lambda_{12} & \lambda_{12} & 0 & 0 & 0\\ 
\lambda_{12}&  \lambda_{11} & \lambda_{12} & 0 & 0 & 0\\
\lambda_{12} &  \lambda_{12} & \lambda_{11} & 0 & 0 & 0\\
0 &  0 & 0 & \frac{\lambda_{11}-\lambda_{12}}{2} & 0 & 0\\
0 &  0 & 0 & 0 & \frac{\lambda_{11}-\lambda_{12}}{2} & 0\\ 
0 &  0 & 0 & 0 & 0 & \frac{\lambda_{11}-\lambda_{12}}{2}\\
\end{bmatrix}
\left[ \hspace{-0.6em} \begin{array}{c} \varepsilon_1 \\ \varepsilon_2 \\ \varepsilon_3 \\ 2\varepsilon_{23} \\ 2\varepsilon_{13} \\ 2\varepsilon_{12} \end{array} \hspace{-0.6em} \right].
\end{equation}

Then, an overall electrical resistivity tensor, $\bm{\rho}_{eff}$, can be defined as the inverse of the conductivity tensor $\bm{\sigma}_{eff}$, given in \eqref{Eq:Electrical_cond}. In the absence of mechanical loading, the electrical resistivity tensor $\boldsymbol{\rho}_{eff}$ takes the form of a scalar matrix with diagonal terms $\rho_0$, i.e.~$\rho_{11}=\rho_{22}=\rho_{33}=\rho_0$ and $\rho_{23}=\rho_{13}=\rho_{12}=0$. Once the composite is subjected to mechanical straining, the components of the resistivity matrix change as follows:
\begin{equation}\label{piezoequation}
\bm{\rho}_{eff}=\rho_{0}\left(\bm{I}+\bm{r}\right).
\end{equation}

The term $\bm{r}$ denotes the tensor of relative change in resistivity and can be related to the mechanical strain tensor $\bm{\varepsilon}$ as $\bm{r}=\bm{\Pi} \colon \bm{\varepsilon}$. Since the piezoresistivity tensor $\bm{\Pi}$ is isotropic~\cite{BURONI2021923}, only two piezoresistivity coefficients ($\lambda_{11}$ and $\lambda_{12}$) suffice to describe it, with the shear coefficient being obtained as $\lambda_{44}=(\lambda_{11}-\lambda_{12})/2$. The closed-form solutions for the effective electrical conductivity and piezoresistivity coefficients presented by Buroni and Garc\'{i}a-Mac\'{i}as~\cite{BURONI2021923} are used in this work.

%%%%%%%%%%%%%%%%%%%%%%%%%%%%%%%%%%%%%%%%%%%%%%%%%%%%
%%%%%%%%%%%%%%%%%%%%%%%%%%%%%%%%%%%%%%%%%%%%%%%%%%%%
%%3.Phase-field
%%%%%%%%%%%%%%%%%%%%%%%%%%%%%%%%%%%%%%%%%%%%%%%%%%%%
%%%%%%%%%%%%%%%%%%%%%%%%%%%%%%%%%%%%%%%%%%%%%%%%%%%%

\section{A phase-field electromechanical model for the fracture of piezoresistive materials}
\label{Section:3}

\subsection{Governing equations}
\label{Section_3_1_governing_eqs}

\begin{figure}[H]
\centering
\includegraphics[scale=1.0]{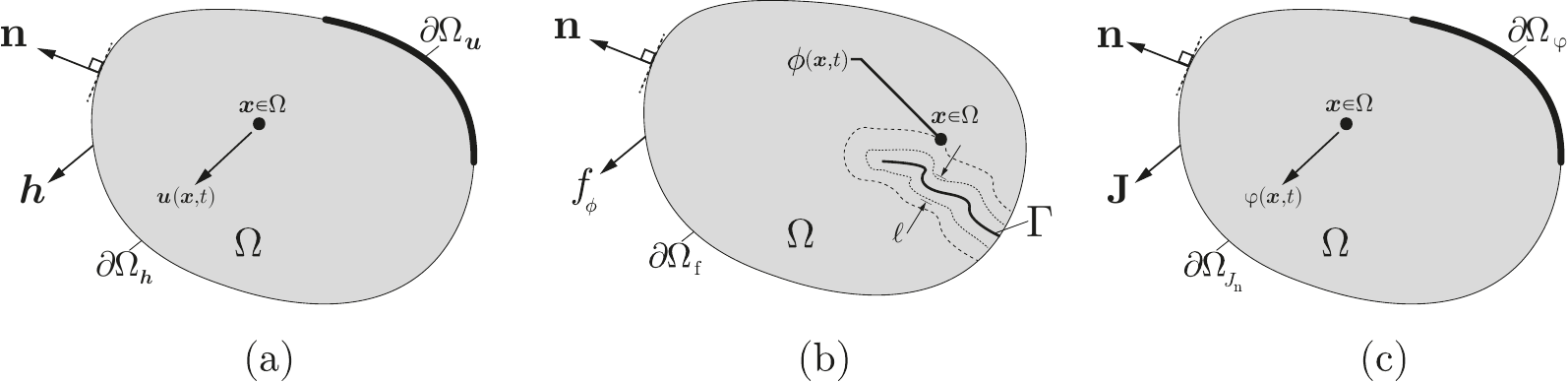}
\caption{Schematic representation of the three-field boundary value problem: (a) deformation, (b) phase-field, and (c) current conservation.}
\label{Fig:Fig_2_1}
\end{figure}

Let us consider a solid domain $\Omega$, whose surface is denoted by $\partial \Omega$ with a normal vector $\mathbf{n}$, as sketched in Fig.~\ref{Fig:Fig_2_1}a. The domain also includes a discontinuous surface $\Gamma$ representing the crack surface. The displacement field and electrical potential are denoted by $\mathbf{u}$ and $\varphi$, respectively. An auxiliary phase-field variable $\phi$ is defined, with values ranging from $\phi=0$ to $\phi=1$, which correspond to the intact and fully broken states of the material, respectively. The phase-field provides a regularisation of the crack surface, whose size is governed by the length scale $\ell$~\cite{Bourdin2000,PTRSA2021}. Regarding the displacement field, the external surface can be decomposed into two parts, a section where the displacements are imposed $\partial \Omega_{u}$, and a second one where the traction boundary conditions $\mathbf{h}$ are imposed $\partial \Omega_{h}$ (Fig.~\ref{Fig:Fig_2_1}a). In addition, an arbitrary crack surface inside the solid $\Gamma$ can be prescribed, and a fracture microtraction $f_{\phi}$ can be prescribed on $\partial \Gamma_{f}$ (Fig.~\ref{Fig:Fig_2_1}b). In turn, a normal electric current flux $J_{n}$ can be prescribed in the boundary $\partial \Omega_{J_{n}}$, whereas the electric potential can be prescribed in the boundary $\partial \Omega_{\varphi}$  (Fig.~\ref{Fig:Fig_2_1}c). In this framework, the principle of virtual work can be formulated, in the absence of body forces, as:
\begin{equation}
 \int_{\Omega} \left(\bm{\sigma} \colon  \delta \bm{\varepsilon}-\textbf{J}\cdot \delta \nabla \varphi  +\omega \cdot \delta \phi +\mathbf{\zeta}\cdot \delta\nabla  \phi\right)\text{d}V=\int_{\partial \Omega} \left( \mathbf{h} \cdot \delta \mathbf{u}+J_{n} \delta \varphi+f_{\phi}\delta \phi \right)\text{d} S,
\end{equation}

\noindent where the operator $\delta$ denotes first-order variations, $\bm{\sigma}$ is the Cauchy stress tensor, $\textbf{J}$ is the flow of electrical current, and  $\omega$ and $\mathbf{\zeta}$ stand for the microstress work quantities conjugate to the phase-field $\phi$ and the phase-field gradient $\nabla \phi$, respectively. Then, applying the Gauss' divergence theorem to the previous expression and using the fundamental lemma of the calculus of variations, one reaches the balance of local forces, which is given by:
\begin{equation} 
    \begin{aligned}
\nabla \cdot \bm{\sigma} &=\mathbf{0}, \quad \\
\nabla \cdot  \mathbf{J} &=0 \quad \text{ in } \quad \Omega, \\
\nabla \cdot \mathbf{\zeta} - \omega &= 0, \quad \\
\end{aligned}
\label{Eq:strong_form}
\end{equation}

\noindent with the natural boundary conditions,
\begin{equation} \label{eq:StrongForm}
    \begin{aligned}
\bm{\sigma} \cdot  \mathbf{n} &= \mathbf{h}\quad \text { on } \quad \partial \Omega_{h},\\
-\mathbf{J} \cdot \mathbf{n}  &= J_{n} \quad \text{ on } \quad \partial \Omega_{ J_{n} }, \\
\mathbf{\zeta} \cdot \mathbf{n} &= f_{\phi} \quad \text { on } \quad \partial \Omega_{f}.\\
\end{aligned}
\end{equation}

\subsection{Constitutive equations}

The deformation-electrical-fracture couplings are as follows. In the first place, the piezoresistivity effect results in an influence of mechanical strains $\bm{\varepsilon}$ on the electrical field $\mathbf{E}$. Also, mechanical straining leads to an increased stored energy $\psi_{0}$ (strain energy density), which is available to nucleate and grow cracks, increasing the magnitude of the phase-field $\phi$. Finally, the presence of cracks impacts the electric conductivity, as captured by degrading the current flux with the phase-field by using an \textit{ad hoc} degradation function. These and other constitutive choices are detailed below. 
\subsubsection{Mechanical deformation}

The strain field under the assumption of small displacements is expressed as:
\begin{equation}
    \bm{\varepsilon}=\frac{1}{2} \left( \nabla \mathbf{u}^{\rm{T}}+\nabla \mathbf{u} \right),
\end{equation}

\noindent and, assuming a linear elastic relationship between the strains and the undamaged stress tensor $\bm{\sigma}_{0}$, the mechanical behaviour of the solid is given by
\begin{equation}
    \bm{\sigma}=h_{1}(\phi)\bm{\sigma}_{0}=h_{1}(\phi)\bm{C} \colon \bm{\varepsilon},
    \label{Eq:Stiff_loss}
\end{equation}

\noindent where $\bm{C}$ is the linear elastic stiffness tensor, and $h_{1}(\phi)$ is a degradation function that relates the phase-field variable with the material stiffness.

\subsubsection{Electrical conductivity}

The relation between the electric field $\mathbf{E}$ and the electric potential $\varphi$ is given by:
\begin{equation}
    \mathbf{E} = -\nabla \varphi,
    \label{Eq:ElectrifieldPotential}
\end{equation}

\noindent while the constitutive equation is given by the linear relation between the conductivity $\bm{\sigma}_{eff}(\bm{\varepsilon})$, which is the inverse of the electrical resistivity $\boldsymbol{\rho}_{eff}$ given in Eq.~(\ref{piezoequation}) (i.e.~$\bm{\sigma}_{eff}(\bm{\varepsilon}) = \boldsymbol{\rho}_{eff}^{-1}$), and the electric current $\mathbf{J}$, which is given by: 
\begin{equation}
    \mathbf{J}= h_{2}(\phi)\bm{\sigma}_{eff}(\bm{\varepsilon}) \mathbf{E}.
    \label{Eq:Condu_loss}
\end{equation}

\noindent Here, $h_{2}(\phi)$ represents a second degradation function that affects the material conductivity, so as to simulate the changes in electrical permeability that take place within cracks. The weak form of the electrical problem can be readily obtained by considering the strong form, Eq. (\ref{Eq:strong_form})b, and making it hold for any admissible $\delta \varphi$. Thus, applying divergence theorem and considering the constitutive definitions (\ref{Eq:ElectrifieldPotential}) and (\ref{Eq:Condu_loss}), one reaches
\begin{equation}
    \int_{\Omega} \left( \delta \nabla \varphi \right)   h_2 (\phi) \bm{\sigma}_{eff}(\bm{\varepsilon})  \nabla \varphi \, \text{d}V = \int_{\partial \Omega_{J_n}} J_{n} \, \text{d}S.
\end{equation}
It is worth noting that the degradation function $h_2$ can modulate sudden changes in electrical conductivity. Thus, while phase-field damage will result in a loss of stiffness and thus high strains, this will not result in a high electric current.

\subsubsection{Phase-field fracture}
\label{Contitutibe_phasefield}

The phase-field fracture model predicts the evolution of cracks as an exchange of stored and fracture energies, building upon the rigorous thermodynamical balance first presented by Griffith \cite{Griffith1920,Francfort1998}. For a cracked solid with strain energy $\Psi (\bm{\varepsilon})$ subjected to a prescribed displacement, Griffith's energy balance can be expressed as the following variation of the total potential energy of the solid $\mathcal{E}$ due to an incremental increase in crack area d$A$:
\begin{equation}\label{eq:Griffith}
\frac{\text{d} \mathcal{E}}{\text{d} A} = \frac{\text{d} \Psi (\bm{\varepsilon})}{\text{d} A} + \frac{\text{d} W_c}{\text{d} A}  = 0,
\end{equation}

\noindent where $W_c$ is the work required to create new surfaces, with the fracture resistance of the solid (or material toughness) being given by $G_c=\text{d}W_c/\text{d}A$. Equation (\ref{eq:Griffith}) can be formulated in a variational form as:
\begin{equation}\label{Eq:Pi}
\mathcal{E} = \int_\Omega \psi \left( \bm{\varepsilon} \right) \text{d} V + \int_\Gamma   G_c \, \text{d} \Gamma \, ,
\end{equation}

\noindent where $\psi$ is the strain energy density of the solid, such that $\Psi=\int \psi \text{d}V$. Then, to make the minimisation of (\ref{Eq:Pi}) computationally tractable, the phase-field paradigm is introduced, whereby an auxiliary variable $\phi$ is used to smear an otherwise discrete interface and track the evolution of that interface. Accordingly, a regularised functional can be formulated as: 
\begin{equation}\label{Eq:Piphi}
\mathcal{E}_\ell = \int_\Omega \left[  h_{1}(\phi) \psi_0 \left( \bm{\varepsilon} \right) + G_c  \left( \frac{\phi^2}{2 \ell} + \frac{\ell}{2} | \nabla \phi |^2 \right) \right] \,  \text{d} V  \, .
\end{equation}

\noindent where $\psi_0$ denotes the strain energy density of the undamaged material, which for an elastic solid reads:
\begin{equation}
    \psi_{0}=\frac{1}{2} \, \bm{\varepsilon}^{T}\colon \bm{C} \colon \bm{\varepsilon}.
\end{equation}

In this work, the regularising term multiplying $G_c$ in (\ref{Eq:Piphi}) is chosen in agreement with the so-called \texttt{AT2} phase-field model \cite{Bourdin2000}. Note also that for piezoresistive materials, the electrical field does not affect the phase-field equation (unlike piezoelectric materials~\cite{MIEHE20101716}). Then, the phase-field constitutive equations can be derived following thermodynamically consistent criteria~\cite{Khalil2022}. Thus, the total potential energy of the solid is given by the sum of the stored and the fracture energy densities as:
\begin{equation}
    \mathcal{W}(\bm{\varepsilon},\phi,\nabla \phi) = h_{1}(\phi)\psi_{0}(\bm{\varepsilon})+ G_{c}\left( \frac{1}{2\ell} \phi^{2}  + \frac{\ell}{2} | \nabla \phi|^{2}\right).
\end{equation}

\noindent The scalar microstress $\omega$ and the vector microstress $\mathbf{\zeta}$ are then derived from the total potential energy as
\begin{equation}
    \omega= \frac{\partial \mathcal{W}}{\partial \phi}= \frac{\partial h_{1}}{\partial \phi} \psi_{0}+G_{c}\frac{\phi}{\ell} , \, \, \, \, \, \, \, \, \, \, \, \text{and} \, \, \, \, \, \, \, \, \, \, \,  \mathbf{\zeta}= \frac{\partial \mathcal{W}}{\partial \nabla \phi}=G_{c} \: \ell \:\nabla \phi.
\end{equation}

\subsection{Degradation functions}

It remains to define the degradation functions $h_{1}(\phi)$ and $h_{2}(\phi)$ introduced in Eqs.~\eqref{Eq:Stiff_loss} and \eqref{Eq:Condu_loss}, respectively. The former describes the loss of stiffness associated with the degradation of material due to damage. For this, we adopt the widely used quadratic function
\begin{equation}
 h_{1}(\phi)=(1-\phi)^{2}.
\end{equation}

On the other side, a degradation function $h_{2}(\phi)$ must be defined to account for the variation in electrical permeability due to cracks. To capture the significant increase in local electrical resistivity observed when the material fractures, we propose the following two-parameter exponential function:
\begin{equation}
h_{2}(\phi,k,n)=\frac{1-\exp{\left(-k(1-\phi)^{n}\right)}}{1-\exp{\left(-k\right)}}.
\label{Eq:df_elec}
\end{equation}

\begin{figure}[H]
\centering
\includegraphics[scale=1]{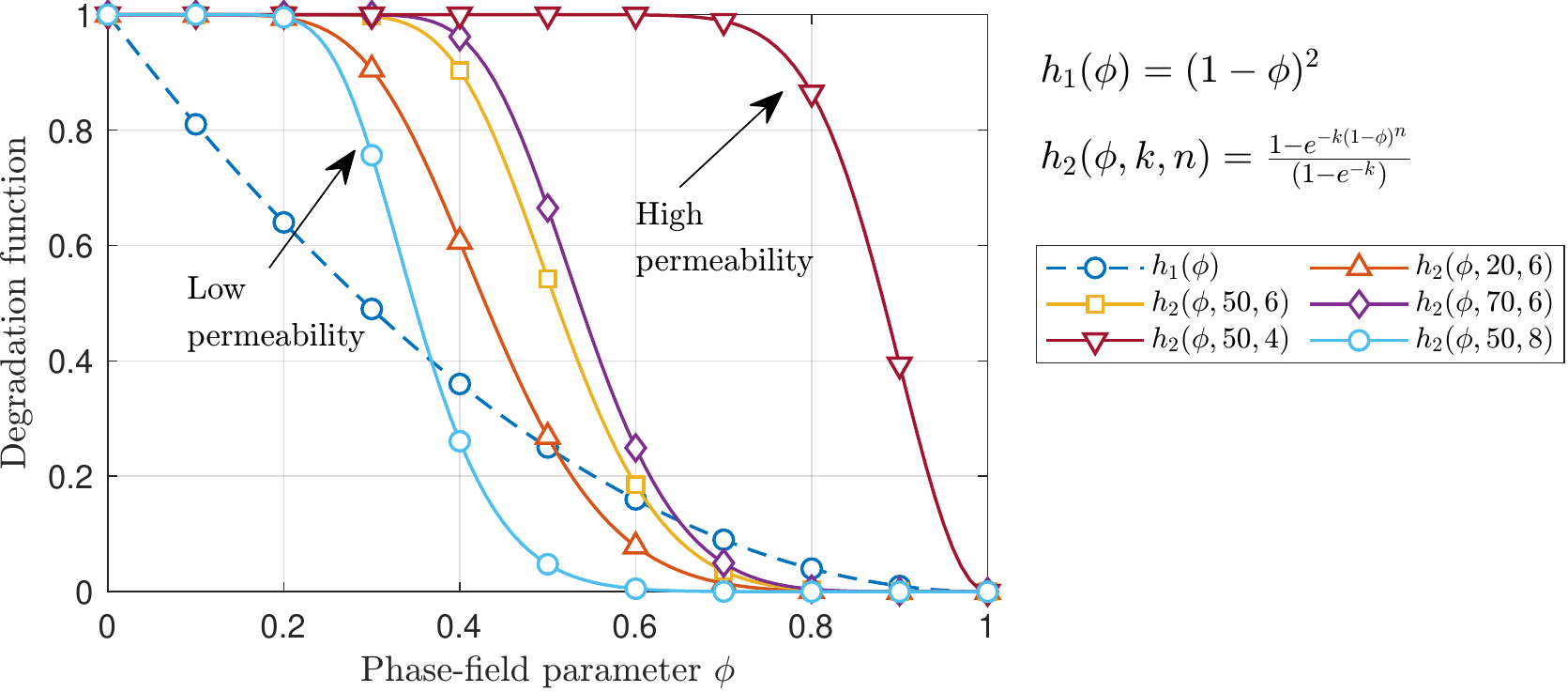}
\caption{Degradation functions employed to interpolate the phase field, $h_{1}(\phi)$, and the electric conductivity, $h_{2}(\phi,k,n)$, with the latter being dependent on the parameters $k$ and $n$.} 
\label{Fig:degra_func}
\end{figure}

The parameters $k$ and $n$ control the shape of the degradation function $h_{2}(\phi,k,n)$, as illustrated in Fig.~\ref{Fig:degra_func}. It can be seen that the shape parameters $k$ and $n$ enable modelling a large variety of degradation functions. For instance, taking $n=6$ and increasing $k$ enables simulating more permeable cracks (higher $h$ values for a given $\phi$). The parameter $n$ instead controls the smoothness of the degradation function, achieving sharp decreases in the electrical conductivity for low values (e.g. $k$=50 and $n$=4), and smooth decreases for large values (e.g. $k$=50 and $n$=8). Finite element predictions will be obtained for various $k$ and $n$ choices to illustrate their influence. As discussed below, the choices $k=50$ and $n=6$ are found to deliver sensible results while ensuring robustness. Thus, they are adopted throughout this work, unless otherwise stated. 

It is also important to note that, for numerical reasons, a small regularization parameter $\epsilon=10^{-7}$ has been added to both $h_{1}(\phi)$ and $h_{2}(\phi,k,n)$ to keep the system of equations well-conditioned.

\subsection{FE implementation}

The finite element (FE) method is chosen to discretise and solve the governing equations provided in Section~\ref{Section_3_1_governing_eqs}. The field variables are the displacement, electric potential and phase-field, which are discretised as:
\begin{equation}
    \mathbf{u}=\sum_{i=1}^{m}\mathbf{N}_{i}\mathbf{u}_{i}, \quad \varphi = \sum_{i=1}^{m}N_{i}\varphi_{i}, \quad \phi =\sum_{i=1}^{m}N_{i}\phi_{i},
\end{equation}

\noindent where $m$ denotes the number of nodes within an element, $N_{i}$ are the shape functions, and $\mathbf{N}_{i}$ corresponds to diagonal matrices with the nodal shape function $N_{i}$ on each component. The strain $\bm{\varepsilon}$, electric field $\mathbf{E}=-\nabla \varphi$, and phase-field gradient $\nabla \phi$ are accordingly interpolated as:
\begin{equation}
    \bm{\varepsilon} =  \sum_{i=1}^{m} \bm{B}_{i}^{\mathbf{u}}\mathbf{u}_{i}, \quad  \mathbf{E}=-\sum_{i=1}^{m}\bm{B}_{i}\varphi, \quad \nabla \phi = \sum_{i=1}^{m}\bm{B}_{i} \phi_{i} ,
\end{equation}

\noindent where $\bm{B}_{i}$ are the spatial derivatives of the shape function and $\bm{B}_{i}^{\mathbf{u}}$ denotes the standard strain-displacement matrices. Using the expression for the momentum equilibrium, phase-field, and electrical current conservation from Eq.~\eqref{Eq:strong_form}, the weak form corresponding to each of the primary fields can be formulated as:
\begin{equation}
    \int_{\Omega}  h_{1}(\phi) \bm{\sigma}_{0} \colon \delta \bm{\varepsilon} \, \text{d}V - \int_{\partial \Omega_{h}} \mathbf{h}\cdot \delta \mathbf{u} \, \text{d} S = 0,
\end{equation}
\begin{equation}
    \int_{\Omega}\left[h_{2}(\phi,k,n) \left(\delta \nabla  \varphi \right) \cdot \bm{\sigma}_{eff}(\bm{\varepsilon})  \nabla \varphi \right] \, \text{d}V - \int_{\partial \Omega_{J_{n}}} \delta \varphi J_{n} \, \text{d}S =0,
\end{equation}
\begin{equation}
    \int_{\Omega} \left[ \frac{\partial h_{1}}{\partial \phi} \delta \phi \psi_{0}+G_{c}\left(\frac{1}{\ell} \phi \delta \phi+ \ell \nabla \phi \cdot \delta \nabla  \phi \right)\right] \, \text{d}V- \int_{\partial \Omega_{f}} f_{\phi}\delta \phi  \, \text{d} S=0. 
\end{equation}

Then, the FE discretization of the residuals can be expressed as:
\begin{equation}
    \mathbf{R}_{i}^{u} =\int_{\Omega} h_{1}(\phi)( \bm{B}_{i}^{u} )^{T}\bm{\sigma}_{0} \, \text{d}V - \int_{\partial \Omega_{h}} \mathbf{N}_{i}^{T} \mathbf{h} \,\text{d}S,
    \label{Eq:Momentum_vector}
\end{equation}
\begin{equation}
    \text{R}_{i}^{\varphi} = \int_{\Omega} \left[ h_{2}(\phi,k,n)\bm{B}_{i}^{T} \bm{\sigma}_{eff}(\bm{\varepsilon})  \nabla \varphi \right] \, \text{d}V - \int_{\partial J_{n}} N_{i}^{T} J_{n} \, \text{d} S,
    \label{Eq:electric_vector}
\end{equation}
\begin{equation}
    \text{R}_{i}^{\phi} = \int_{\Omega} \left[ G_{c} \left( \frac{1}{\ell} N_{i} \phi + \ell \bm{B}_{i}^{T} \nabla \phi \right)+\frac{\partial h_{1}}{\partial \phi} N_{i} \mathcal{H} \right] \, \text{d} V - \int_{\partial \Omega_{f}} N_{i} f_{\phi} \, \text{d} S,
    \label{Eq:phasefield_vector}
\end{equation}

\noindent in which we adopt the so-called history variable $\mathcal{H}$ \cite{MIEHE20102765} to ensure damage irreversibility, such that $\mathcal{H}=\text{max}_{t \in [0,t_t]} \psi (t) $ for a time $t$ within a total time $t_t$. Finally, the corresponding stiffness matrices can be stated as:
\begin{equation}
    \bm{K}_{ij}^{u} = \frac{\partial \mathbf{R}_{i}^{u}}{\partial \mathbf{u}_{j}} = \int_{\Omega}h_{1}(\phi) ( \bm{B}_{i}^{u} )^{T} \bm{C}  \bm{B}_{j}^{u} \text{d} V,
    \label{Eq:Momentum_matrix}
\end{equation}
\begin{equation}
    \bm{K}_{ij}^{\varphi} = \frac{\partial \mathbf{R}_{i}^{u}}{\partial \varphi_{j}}=\int_{\Omega} h_{2}(\phi,k,n) ( \bm{B}_{i}) ^{T} \bm{\sigma}_{eff}(\bm{\varepsilon})  \bm{B}_{j} \text{d}V,
    \label{Eq:electric_matrix}
\end{equation}
\begin{equation}
    \bm{K}_{ij}^{\phi} = \frac{\partial \mathbf{R}_{i}^{\phi}}{\partial \phi_{j}} = \int_{\Omega} \left[ \left( 2 \mathcal{H} + \frac{G_{c}}{\ell}  \right) N_{i} N_{j} + G_{c} \ell \bm{B}_{i}^{T} \bm{B}_{j} \right] \text{d} V.
    \label{Eq:phasefield_matrix}
\end{equation}

\noindent And thus the deformation-electrical-damage FE system can be expressed as
\begin{equation}
\left\{\begin{array}{l}
\mathbf{u} \\
\mathbf{\varphi}\\
\mathbf{\phi}
\end{array}\right\}_{t+\Delta t}=\left\{\begin{array}{l}
\mathbf{u} \\
\mathbf{\varphi}\\
\mathbf{\phi}
\end{array}\right\}_{t}-\left[\begin{array}{ccc}
\bm{K}^{u} & 0 & 0 \\
0 & \bm{K}^{\varphi} & 0\\
0 &0 & \bm{K}^{\phi}
\end{array}\right]_{t}^{-1}\left\{\begin{array}{l}
\mathbf{R}^{\mathbf{u}} \\
\mathbf{R}^{\varphi} \\
\mathbf{R}^{\phi}
\end{array}\right\}_{t}.
\label{Eq:Matricial_scheme}
\end{equation}
The system given in Eq.~\eqref{Eq:Matricial_scheme} is fully coupled. Mechanical deformation influences both the phase-field variable and the electrical potential by means of the strain energy density and the piezoresistive properties, respectively. In addition, the phase-field degrades the stiffness of the solid and the electrical conductivity by means of the degradation functions described in Section~\ref{Contitutibe_phasefield}. These couplings are taken care of by using a monolithic scheme that ensures unconditional stability. Robustness and efficiency within a monolithic solution scheme are achieved by approximating the stiffness matrix in Eq.~\eqref{Eq:Matricial_scheme} by means of quasi-Newton methods. Specifically, the Broyden-Fletcher-Goldfarb-Shanno (BFGS) algorithm is used, as it has proven to lead to efficient and robust monolithic phase-field fracture implementations \cite{KRISTENSEN2020102446,WU2020112704}. The deformation-electrical-fracture model is implemented into the finite element package \texttt{Abaqus} as a user-element (\texttt{UEL}) subroutine, which is openly shared \footnote{The \texttt{UEL} subroutine developed can be found in \url{www.imperial.ac.uk/mechanics-materials/codes} and \url{www.github.com/L-Quinteros}}. For generality, the finite element implementation is carried out in a linear brick element with 8 nodes, full integration and 5 degrees-of-freedom (DOFs) per node ($u_x$, $u_y$, $u_z$, $\varphi$, $\phi$). However, the implementation is also particularised to 2D scenarios is some of the case studies considered below.

%%%%%%%%%%%%%%%%%%%%%%%%%%%%%%%%%%%%%%%%%%%%%%%%%%%%
%%%%%%%%%%%%%%%%%%%%%%%%%%%%%%%%%%%%%%%%%%%%%%%%%%%%
%%%%%%%%%%%%%%%%%%%%%%%%%%%%%%%%%%%%%%%%%%%%%%%%%%%%
%%4.Results
%%%%%%%%%%%%%%%%%%%%%%%%%%%%%%%%%%%%%%%%%%%%%%%%%%%%
%%%%%%%%%%%%%%%%%%%%%%%%%%%%%%%%%%%%%%%%%%%%%%%%%%%%
\section{Results}
\label{Section:4-Results}

In this section, we evaluate the performance of the model in simulating crack initiation and propagation and their effects upon the electromechanical response of epoxy composites doped with multi-walled CNTs (MWCNTs). The electromechanical properties of the constituents of the considered composites have been taken from the literature and are collected in Table~\ref{fittedparam}. Parametric analyses to evaluate the effect of the filler volume fraction and CNT aspect ratio are firstly reported in Section~\ref{Section4:props}. Then, five case studies are presented. The first case study validates the proposed formulation against the experimental data reported by Esmaeili~\textit{et al.}~\cite{esmaeili2020} (Section \ref{Section4:case4}). Further insight is gained by considering three case studies involving plane boundary value problems with different configurations of initial defects (Sections \ref{Section4:case1} to \ref{Section4:case3}). Finally, three-dimensional crack nucleation and growth is investigated in Section~\ref{Section4:case5}. 
\begin{center}

\begin{table}[h]
\small

\newcommand\Tstrut{\rule{0pt}{0.3cm}}         % = `top' strut									
\newcommand\Bstrut{\rule[-0.15cm]{0pt}{0pt}}   % = `bottom' strut								
 \footnotesize												
  \caption{Material parameters and micromechanical variables adopted. The values chosen correspond to those of MWCNT/epoxy composites, and are taken from Refs. ~\cite{menna2016effect,Garcia-Macias2018b}.}								
 \vspace{0.1cm}												
 \centering	
\resizebox{\textwidth}{!}{
\begin{tabular}{llllll}		
	\hline 	
Name & Symbol & Value & Name & Symbol & Value \Tstrut\\
  \hline
Volume fraction & $f_{p}$ & 1\% & Length of MWCNT          & $L_{cnt}$	&	\SI{3.21}{\mu \meter} \\ 
Outer diameter of MWCNT & $D_{cnt}$	&	10.35	nm			& Cut-off distance for tunnelling effects & $d_c$	&	0.22	nm \\
Height of the potential barrier &$\lambda$	&	0.69	eV	& Elastic modulus of CNT&	$E_{cnt}$	&	700	GPa	 \\
Elastic modulus of epoxy&$E_m$	&	2.5	GPa	& Electrical conductivity of MWCNT & $\sigma_c$	&	100	S/m	\\
Electrical conductivity of epoxy & $\sigma_m$	&	1.036E-10	S/m	& Possion's ratio	of MWCNT   & $\nu_{cnt}$	&	0.3\\
Possion's ratio of epoxy & $\nu_m$	&	0.28		& Interphase thickness & $t$	&	31.00	nm \\
Elastic modulus of interphase &	$E_i$	&	2.17	 GPa	&
%Density of MWCNT & $\rho_{cnt}$	&	1.42	g/cm$^3$	\\
%Density of epoxy	&	$\rho_m$	&	1.12	g/cm$^3$	\\
Strength of CNT & $\sigma_{cnt}$ & 35 GPa \\
Interfacial shear strength & $\tau_{cnt}$ & 47 MPa  &
Fracture energy of pristine epoxy & $G_{ce}$ & 133 J/m$^2$ \\
Experimental orientation limit angle & $A$& 0.083 &
Minimum CNT orientation angle & $\theta_{min}$ & 0 \\
Maximum CNT orientation angle & $\theta_{max}$ &$\pi/2$ & & & \\ 
\hline 												
 \end{tabular}	
 }
\label{fittedparam}												
 \end{table}	

\end{center}

\subsection{Estimation of constitutive properties}
\label{Section4:props}

\begin{figure}[H]
\centering
\includegraphics[scale=0.9]{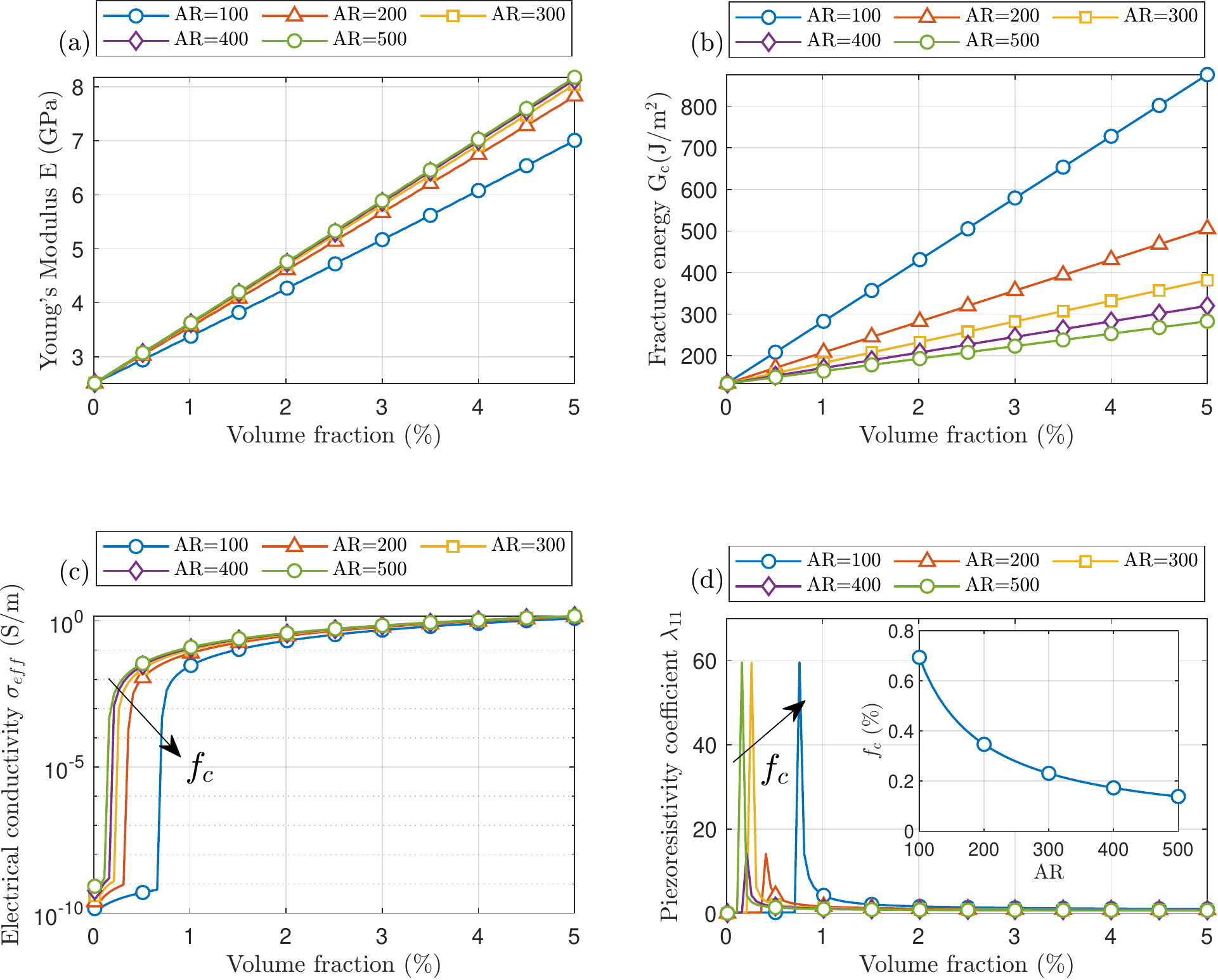}
\caption{Electromechanical properties of epoxy/MWCNT composites estimated by the presented micromechanical framework, including:  (a) the elastic modulus, (b) fracture energy, (c) electrical conductivity, and (d) the piezoresistivity coefficient $\lambda_{11}$, with the insert representing the variation of the percolation threshold $f_c$ as a function of the filler aspect ratio (AR).} 
\label{Fig:contitutive_prop}
\end{figure}

The formulation previously presented in Section~\ref{section 2: Modelling} is adopted to estimate the elastic moduli, critical energy release rate, electrical conductivity, and linear piezoresistivity constant $\lambda_{11}$ of epoxy/MWCNT composites for a wide range of filler volume fractions and aspect ratios $\text{AR}=L_{cnt}/D_{cnt}$, as reported in Figure~\ref{Fig:contitutive_prop}. The results show that the effective elastic modulus and the fracture energy follow a linear fashion, whereas the electrical conductivity and the piezoresistivity coefficient $\lambda_{11}$ exhibit non-linear behaviour. The elastic modulus and the fracture energy increase with the volume fraction, while the opposite behaviour is observed for increasing aspect ratios (ARs). It is noted in Fig.~\ref{Fig:contitutive_prop}a that the elastic modulus shows a fast convergence rate for increasing ARs (no significant enhancements are found for ARs above 300), while a slower convergence is observed for the fracture energy $G_c$. The fracture energy is mainly governed by the pull-out mechanism from Eq.~\eqref{Eq:W_en} showing a critical embedded length dependency which, in turn, has a diameter dependency, increasing this value at low aspect ratios. A very different trend is observed for the electrical conductivity and the piezoresistivity coefficient $\lambda_{11}$. Firstly, they show almost no senitivity for low volume fractions, where the electron hopping mechanism dominates. However, a significant rise is observed when the CNT volume fraction reaches the percolation threshold, representing the onset of the networking mechanism. It is also interesting to note in the insert in Fig.~\ref{Fig:contitutive_prop}d the variation of the percolation threshold $f_{c}$ as a function of the filler aspect ratio. This result evidences the fact that fillers with large aspect ratio favour the development of conductive networks, which in turn manifests as lower percolation thresholds.

\subsection{Experimental validation}
\label{Section4:case4}

\begin{figure}[H]
\centering
\includegraphics[scale=0.92]{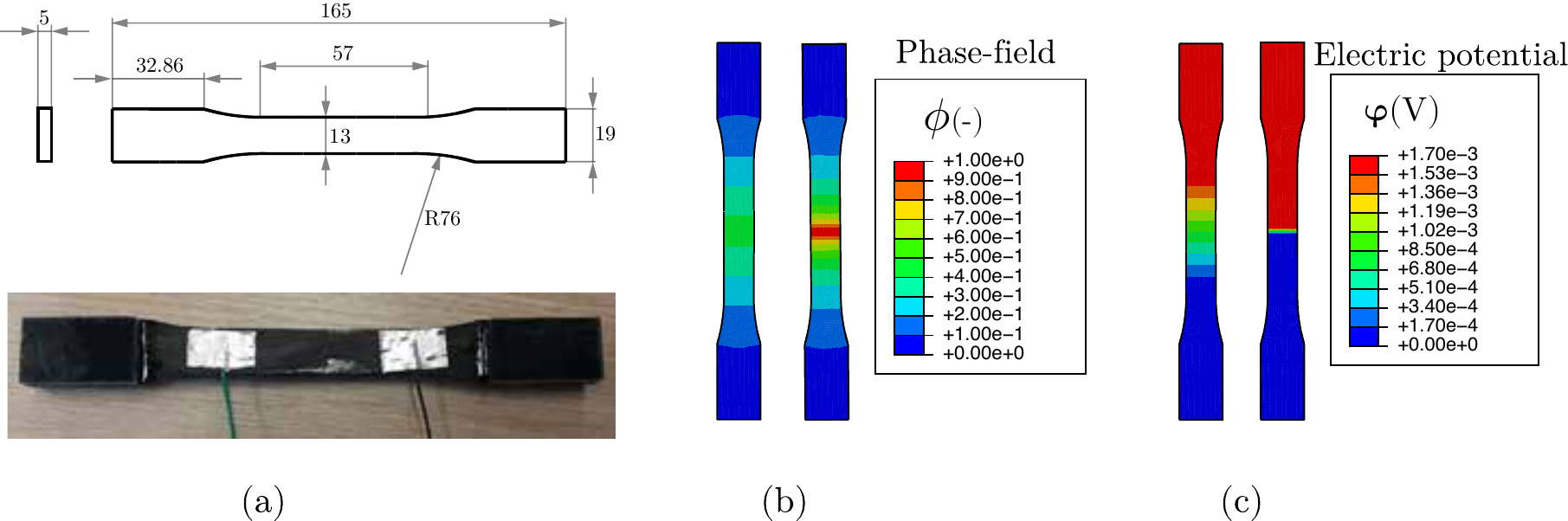}
\caption{Experimental validation: (a) test dimensions (units in mm) and configuration, contours of (b) the phase-field variable, and (c) the electrical potential, before ($\varepsilon_{1} = 0.0123$) and after ($\varepsilon_{1} = 0.0124$) cracking.} 
\label{Fig:Fig_exp_1}
\end{figure}

\begin{figure}[H]
\centering
\includegraphics[scale=0.9]{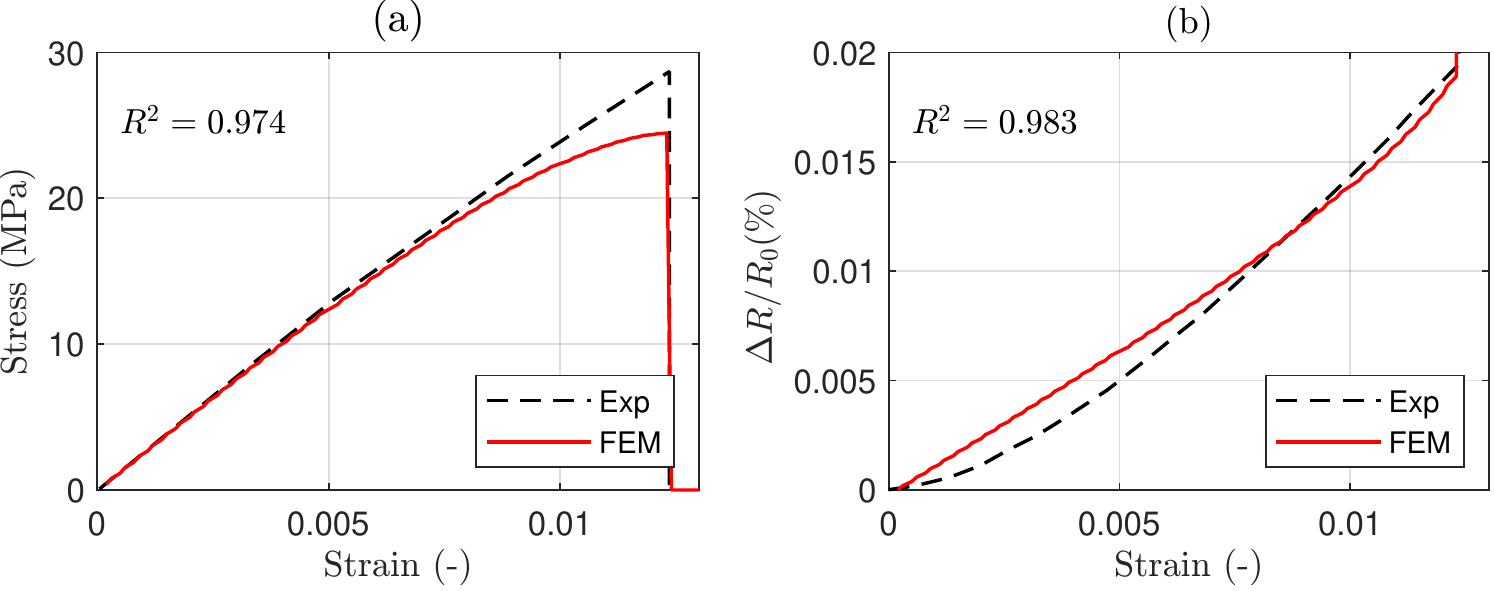}
\caption{Comparison between experimental \cite{esmaeili2020} and numerical predictions of DGBA/DWCNT composite behaviour: (a) stress - strain curve, and (b) relative variation of the electrical resistance versus externally applied strain.} 
\label{Fig:Fig_exp}
\end{figure}

The following case study presents the validation of the proposed model using the experimental data presented in the work of Esmaeili~\textit{et al.}~\cite{esmaeili2020}. Those authors reported the electromechanical characterization of dog-bone samples made of bisphenol A diglycidyl ether (DGBEA) epoxy doped with single and double walled CNTs (SWCNTs-DWCNTs) subjected to tension until failure. The specimen dimensions are shown in Fig.~\ref{Fig:Fig_exp_1}a. In the numerical model, the sample is discretised using a total of 197,120 DOFs, with the characteristic element length in the relevant regions being 7 times smaller than the phase-field length scale ($\ell = 0.0012$ mm). The structure is subjected to vertical displacements on its top edge, while the bottom edge is pinned. The electrodes are located at 5 cm from each other and subjected to a potential difference of 1.7 mV applied in a $13 \times 5$ $\text{mm}^{2}$ area. The mass fraction of CNTs is 0.5 wt\% and the remaining micromechanical parameters in Table~\ref{Tab:val_sc4} were obtained by curve fitting with the experimental results, using typical values reported in the literature for DGBA epoxy and CNTs. The electrical resistance reported in the experiments  between the electrodes was about 8500 $\Omega$, while the resistance predicted by the numerical model is 8485 $\Omega$, demonstrating the effectiveness of this model regarding the unstrained state of the composite. The phase-field and electric potential contour plots are reported in Figs.~\ref{Fig:Fig_exp_1}b and c, respectively. The stress-strain and the relative resistance-strain curves are shown in Figs.~\ref{Fig:Fig_exp}a and b, respectively. Both curves show good agreements with the experimental data. It is noted that the numerical stress-strain curve exhibits a slight decrease at high tensile strains unlike the experimental results, which exhibit a clear linear tendency. These differences are ascribed to the assumed quadratic degradation function of the stiffness and the use of the so-called \texttt{AT2} model, which lacks a purely elastic domain. The results in Figure~\ref{Fig:Fig_exp}b evidence the presence of a slightly more marked non-linear behavior in the experimental data compared to the numerical simulation. In the literature, the presence on non-linearities in the strain sensing curves has been identified to be driven by strain-induced variations in the contribution by the electron hopping mechanism. In the context of the implemented micromechanics approach in Section~\ref{Section 2.3: Electrical properties}, this may indicate some limitations in the theoretical definition of the strain dependency of the resistivity properties related to the quantum tunnelling effects (refer to e.g.~\cite{GARCIAMACIAS2017195} for further discussion in this regard). Nonetheless, the accuracy of the adopted micromechanics approach is considered sufficient for the aim of the present work, especially for the analysis of strain sensing applications where non-linear effects in the piezoresitive CNT-based composites are very limited.

\begin{table}[h]	
\small

\newcommand\Tstrut{\rule{0pt}{0.3cm}}         % = `top' strut									
\newcommand\Bstrut{\rule[-0.15cm]{0pt}{0pt}}   % = `bottom' strut								
 \footnotesize												
  \caption{Micromechanical variables adopted for the experimental validation against tests on a DGBA/DWCNT composite. The values used lie within the range reported for DGBA/DWCNT composites in the literature \cite{menna2016effect,Garcia-Macias2018b,esmaeili2020}.}		

 \vspace{0.1cm}												
 \centering		
\resizebox{\textwidth}{!}{
 \begin{tabular}{llllll} 	
	\hline 	
Name & Symbol & Value & Name & Symbol & Value \Tstrut\\
  \hline
Mass fraction & $w_{p}$& 0.5 \% 
& Length of CNT          & $L_{cnt}$	&	5.39	$\mu$m \\
Outer diameter of CNT & $D_{cnt}$	&	1.203	nm			& Cut-off distance for tunnelling effects & $d_c$	&	2.739	nm \\
Height of the potential barrier &$\lambda$	&	1.93	eV	& Elastic modulus of CNT&	$E_{cnt}$	&	950	GPa	 \\
Elastic modulus of epoxy&$E_m$	&	2.79	GPa	 & Electrical conductivity of CNT & $\sigma_c$	&	764.91	S/m	\\
Electrical conductivity of epoxy & $\sigma_m$	&	1.00E-12	S/m	& Possion's ratio	of CNT   & $\nu_{cnt}$	&	0.3 \\
Possion's ratio of epoxy & $\nu_m$	&	0.285		& Interphase thickness & $t$	&	31.00	nm \\
Elastic modulus of interphase &	$E_i$	&	2.24	 GPa	& Density of CNT & $\rho_{cnt}$	&	1.35	g/cm$^3$	\\
Interfacial shear strength & $\tau_{cnt}$ & 47 MPa & Fracture energy of pristine epoxy & $G_{ce}$ & 220 J/m$^2$ \\
Experimental orientation limit angle & $A$& 0.083 &
Minimum CNT orientation angle & $\theta_{min}$ & 0 \\
Maximum CNT orientation angle & $\theta_{max}$ &$\pi/2$ & Density of epoxy	&	$\rho_m$	&	1.15	g/cm$^3$ \\ 
Strength of CNT & $\sigma_{cnt}$ & 120 GPa & & & \\
\hline 												
 \end{tabular}
 }
\label{Tab:val_sc4}	

 \end{table}

\subsection{Mixed-mode fracture of a thin plate containing an initial crack}
\label{Section4:case1}

\begin{figure}[H]
\centering
\includegraphics[scale=0.85]{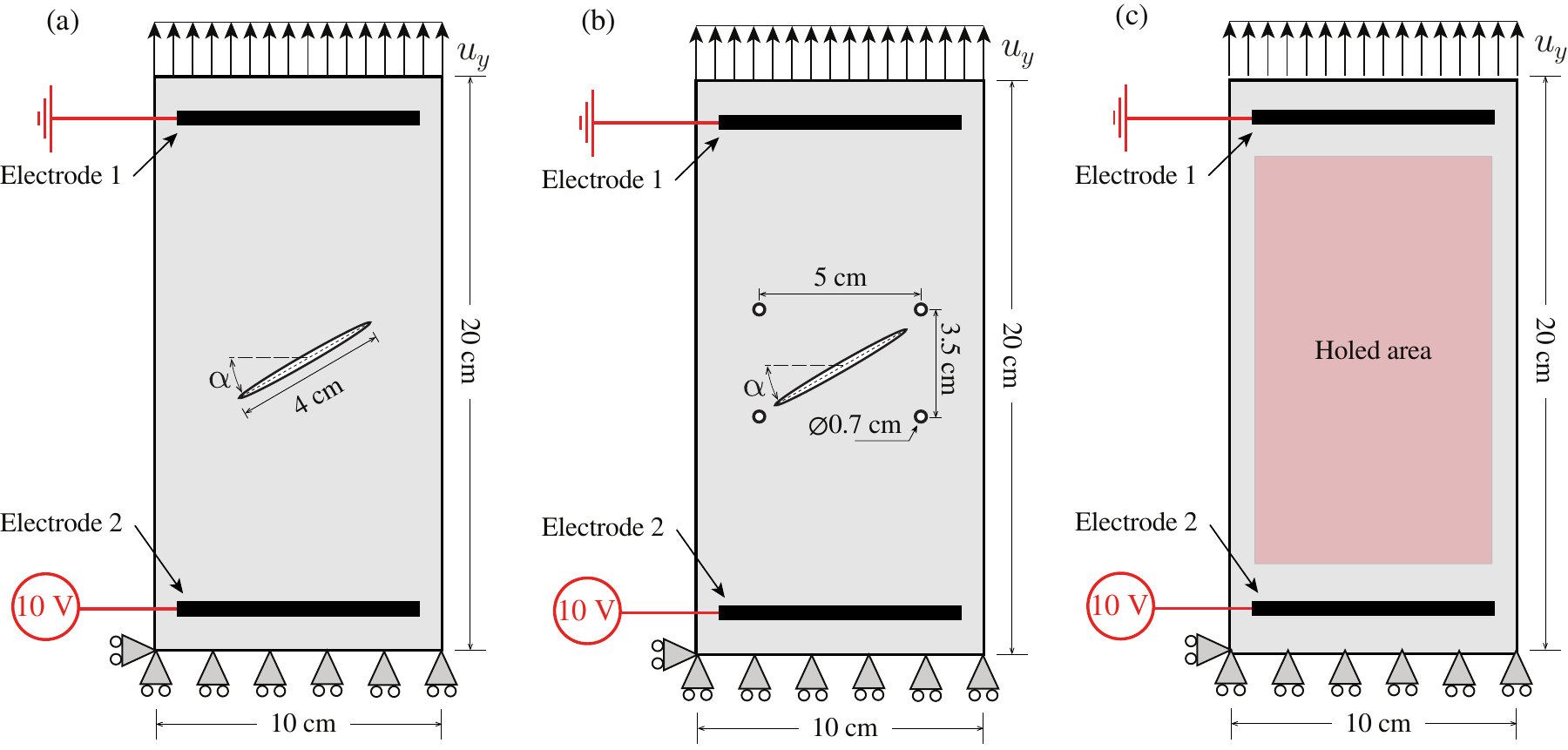}
\caption{Schematic of the plate geometry and boundary conditions of three plane case studies consisting of: (a) a plate containing an inclined crack, (b) a plate containing an inclined crack and multiple holes, and (c) a plate containing a random distribution of holes. All are under the same electric and displacement boundary conditions.}
\label{Fig:model}
\end{figure}

The second case study involves a 10 cm by 20 cm notched plate with a thickness of 0.5 cm subjected to vertical displacements on the top edge and pinned on the bottom edge. The notch is imposed geometrically, and the electric potential is imposed using two electrodes as shown in Fig.~\ref{Fig:model}a. The upper electrode is grounded while a differential potential of 10 V is applied at the bottom edge. In the following analyses, a volume fraction of $f_{p}=1 \%$, and the degradation parameters $k=50$ and $n=6$ are considered. In this and all remaining case studies, the material properties employed are those provided in Table \ref{fittedparam}. In this regard, it is worth noticing that the CNT aspect ratio assumed $L_{cnt}/D_{cnt}=310$, lies within the regime where fibre pull-out dominates over the fibre rupture mechanism, as per the sensitivity analyses conducted in Ref. \cite{QUINTEROS2022109788}.

\begin{figure}[H]
\centering
\includegraphics[scale=0.95]{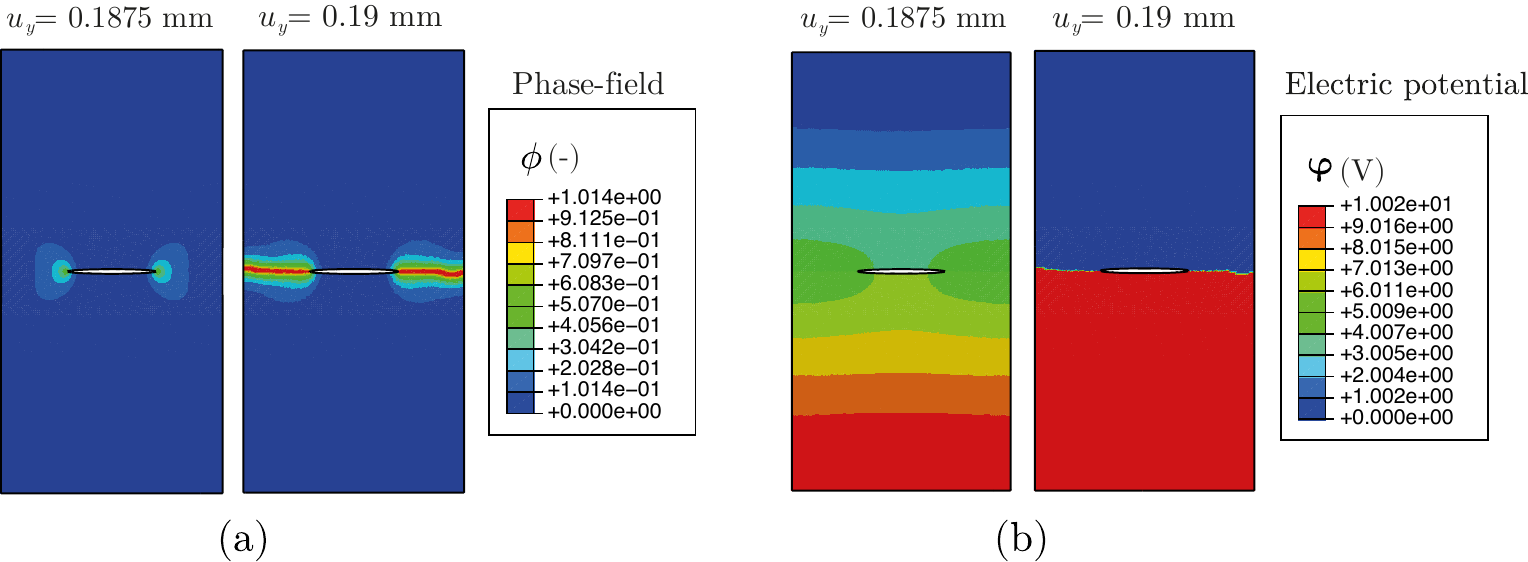}
\caption{Mixed-mode fracture of a plate with an initial crack. Contours of (a) the phase-field variable $\phi$, and (b) the electric potential $\varphi$, for two values of the remote displacement (before and after full fracture).}
\label{Fig:Study_1}
\end{figure}

The domain is discretised with approximately 100,000 DOFs, with the characteristic element size in the potential crack growth regions being equal to 0.002 mm, three times smaller than the phase-field length scale $\ell$. Figure~\ref{Fig:Study_1} shows the contour plot of the phase-field variable and the electrical potential. Figure~\ref{Fig:Study_1}a shows the evolution of the phase-field variable $\phi$ at three different instants with imposed displacements $u_{y}=0.1975$ \rm{mm}, $u_{y}=0.1875$ \rm{mm}, and $u_{y}=0.19$ \rm{mm}. The phase-field value increases around the crack tip and then the crack is shown to propagate horizontally. Before complete failure, the phase-field barely affects the electrical potential and only linear variations induced by piezoresistivity are observed. However, once the plate cracks, sudden decreases in the electrical current flowing through the electrodes are noted. This is evidenced in the contour plot of electric potential in Fig.~\ref{Fig:Study_1}(b), in which, once the crack develops all throughout the specimen, two distinct zones are noted with electric potentials of 0 and 10 \rm{V} due to the very low electrical permeability (or very high electrical resistivity) of the cracked domain. 

\begin{figure}[H]
\centering
\includegraphics[scale=0.83]{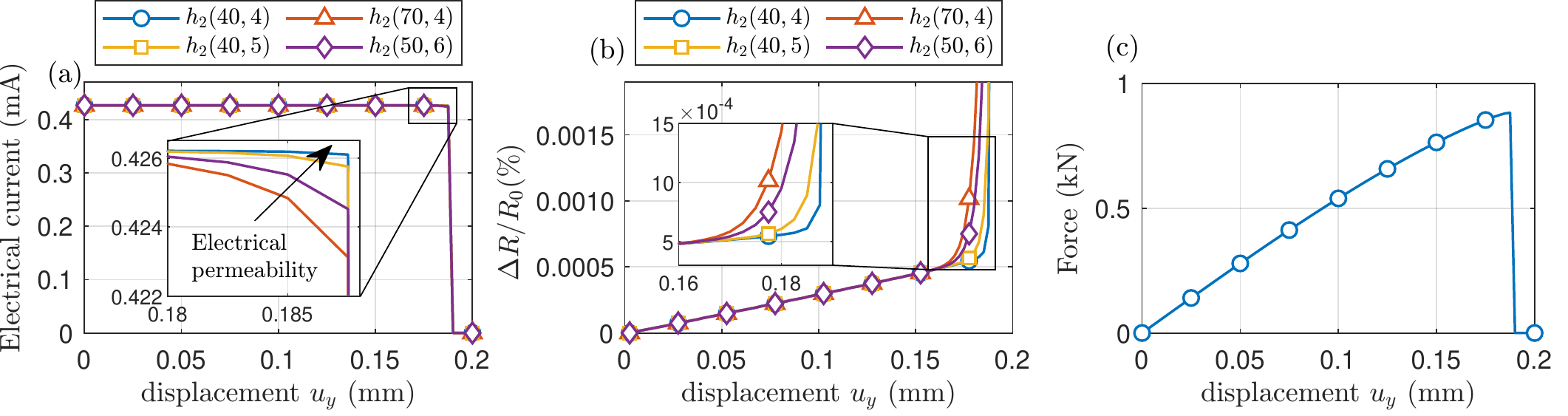}
\caption{Effect of the degradation function $h_{2}(\phi,k,n)$ on: (a) the electrical current, (b) relative resistance $\Delta R/R_{0}$, and (c) its corresponding load displacement curve.} 
\label{Fig:case_1_h2}
\end{figure}

The impact of the shape parameters $k$ and $n$ of the degradation function $h_{2}(\phi,k,n)$ on the electrical current flowing between the electrodes is investigated in Fig.~\ref{Fig:case_1_h2}a. Note that all the curves correspond to the same load-displacement curve presented in Fig.~\ref{Fig:case_1_h2}c. For all cases, the results shown in Fig.~\ref{Fig:case_1_h2}a show that the curves slightly decrease until the plate cracks, provoking a steep descent of the conductivity. It can be observed that the slope of the electrical conductivity changes noticeably depending on $k$ and $n$. In practice, the degradation function $h_{2}(\phi,k,n)$ may be calibrated by fitting experimental data. Figure~\ref{Fig:case_1_h2}b presents the relative variation of the electrical resistance, which is calculated using the unstrained resistance $R_{0}$ and the instantaneous electrical resistance $R$ as $(R-R_{0})/ R_{0} $. In this case, the results show a dramatic increase of the electrical resistance as soon as the displacement reaches the fracture displacement.

\begin{figure}[H]
\centering
\includegraphics[scale=0.83]{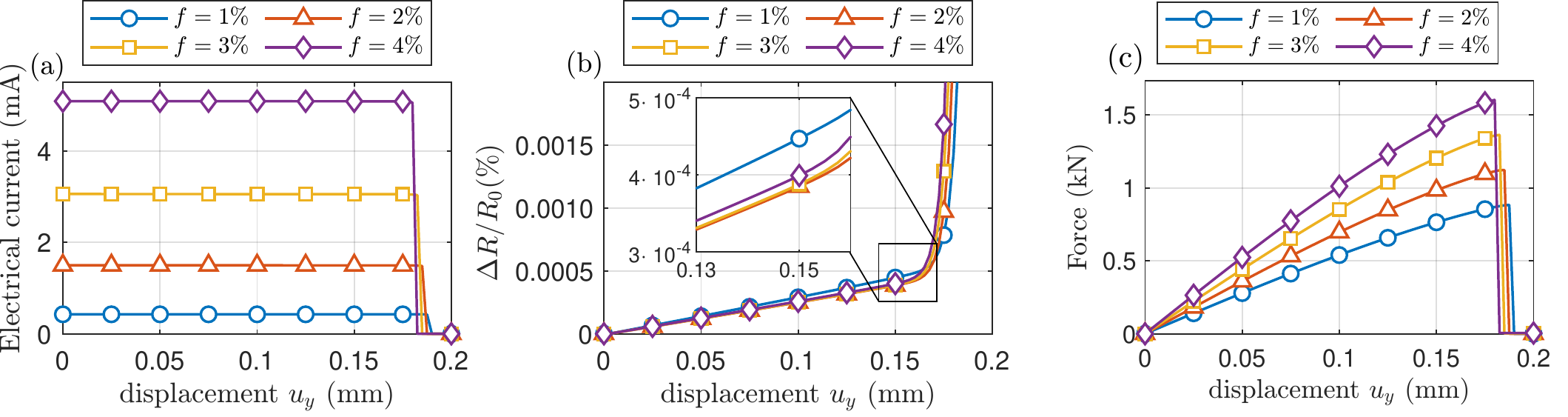}
\caption{Effects of the filler volume fraction $f_{p}$ upon: (a) the electrical current-displacement curve, (b) the relative resistance $\Delta R/R_{0}$, and (c) the load-displacement curve.} 
\label{Fig:case_1_fp}
\end{figure}

Figure~\ref{Fig:case_1_fp} shows the effects of the CNT volume fraction on the electrical current flowing between the electrodes, the corresponding relative variation of electrical resistance, and the load-displacement curve. In Fig.~\ref{Fig:case_1_fp}a, it can be seen that the consideration of higher CNT volume fractions increases the electrical current, as a result of the enhanced effective conductivity of the composite. In this figure, two distinct regimes of behaviour are clearly noticeable: before and after crack initiation. The first regime is dominated by linear decreases driven by the piezoresistive property of the composite. Instead, once the crack initiates, the electrical conduction through the specimen is dominated by the permeability of the crack. For instance, in the case of epoxy doped with $f_{p}= 4\%$ CNTs at the beginning of the displacement load process, the electrical current is 5.8782 \rm{mA} and it decreases to 5.8604 \rm{mA}, right before fracture. Finally, the electrical current goes to zero when the crack crosses the whole cross-section of the specimen, indicating the complete interruption of the current flow. Figure~\ref{Fig:case_1_fp}b shows the relative resistance versus the imposed displacement $u_{y}$, for different filler volume fractions. It is observed in this figure that the addition of higher volume fractions leads to higher piezoresistivity coefficients, as indicated by the larger slopes of the first linear range. The addition of higher concentrations of CNTs also enhances the effective mechanical properties of the composite, as evidenced by the increases of the slopes of the load-displacement curves of Fig.~\ref{Fig:case_1_fp}c. Note that the addition of CNTs diminishes the displacement $u_{y}$ in which the plate breaks, which can be readily explained from an energetic standpoint. As the improvement in the elastic modulus induced by the addition of CNTs increases, the area under the load-displacement curve raises and, consequently, the fracture displacement decreases. Note in Fig.~\ref{Fig:contitutive_prop}b that the critical energy release rate experiences comparatively smaller raises in magnitude for increasing filler contents, relative to the elastic modulus.

\begin{figure}[H]
\centering
\includegraphics[scale=0.83]{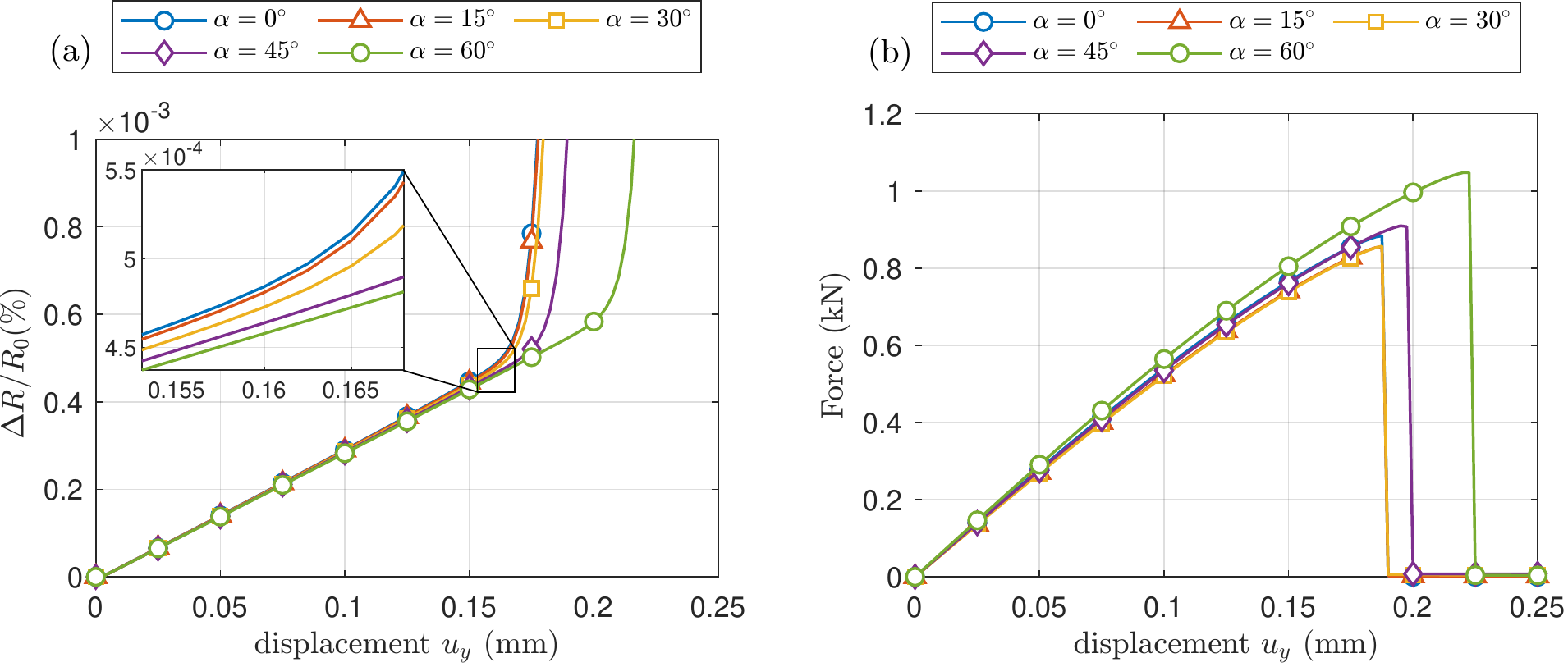}
\caption{Effect of the crack angle $\alpha$ on: (a) the relative resistance-displacement curves, and (b) the load-displacement curves.} 
\label{Fig:case_1_angle}
\end{figure}

The effect of the crack inclination angle $\alpha$ is studied in Fig.~\ref{Fig:case_1_angle}. The relative resistance can be observed in Fig.~\ref{Fig:case_1_angle}a as a function of the displacement. It is noted in Fig.~\ref{Fig:case_1_angle}b that increasing the notch angle $\alpha$ raises the ultimate load capacity and the critical displacement at failure. This is due to the reduction of the stress concentration at the crack tips as the projection of the crack surface with respect to the direction of the imposed displacement decreases (a move from mode I fracture to mixed-mode conditions). This effect is also evident in Fig.~\ref{Fig:case_1_angle}a in terms of relative variations of the electrical resistance of the specimen. As the notch angle increases, the degradation of the electrical conductivity induced by damage appears for higher imposed displacements. These results demonstrate the usefulness of electrical resistivity measurements to infer the appearance and geometrical properties of crack-like defects.

\subsection{Fracture of a thin plate containing an initial crack and nearby circular defects}
\label{Section4:case2}

\begin{figure}[H]
\centering
\includegraphics[scale=0.9]{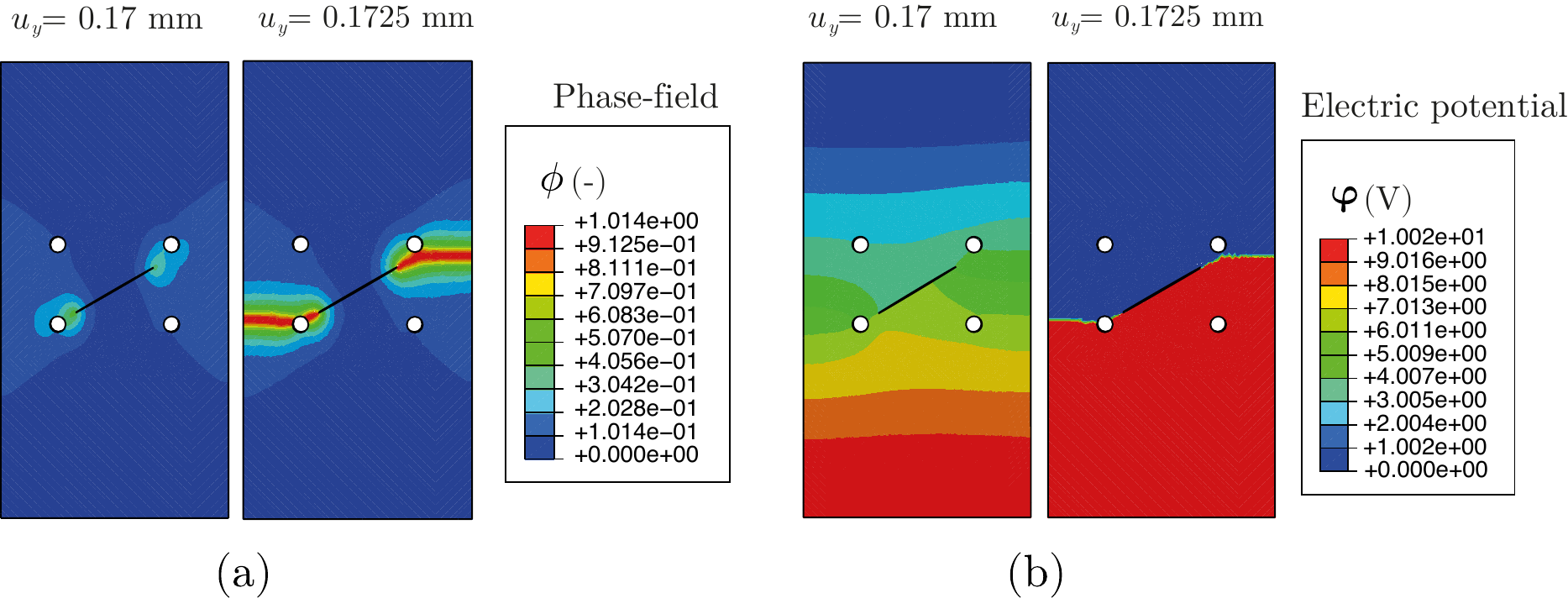}
\caption{Fracture of a plate containing an initial crack and nearby circular defects. Contours of (a) the phase-field variable $\phi$, and (b) the electric potential $\varphi$, for different values of the remote displacement. Representative results obtained for a crack inclination angle of $\alpha=30^{\circ{}}$.}
\label{Fig:case_2}
\end{figure}

This case study considers the exact same geometry as in the previous one but with the addition of four holes around the notch. The material parameters reported in Table~\ref{fittedparam} are adopted herein. A total of approximately 90,000 DOFs are used to discretise the model, with the phase-field length scale ($\ell = 0.006$ mm), being more than three times larger than the characteristic element size. A plate with an inclined crack of 30 degrees is studied first. Figure~\ref{Fig:case_2}a shows the phase-field $\phi$ at two stages, where the displacement at the top edge during the loading process is $u_{y}=0.17$ \rm{mm} and when it equals $u_{y}=0.1725$ \rm{mm}. Figure~\ref{Fig:case_2}b instead shows the electrical potential $\varphi$ during the fracture process. It is noted in Fig.~\ref{Fig:case_2}a that the breakage of the plate develops in two phases. Firstly, the crack propagates from the notch to the holes closest to the notch along the diagonal, to then propagate until crossing the whole plate. Finally, once the plate is fully cracked, the electrical flow between the electrodes is interrupted as evidenced by the contour plot in Fig.~\ref{Fig:case_2}b for $u_{y}=0.1725$ mm where the crack concentrates the voltage drop from 10 to 0 V.

\begin{figure}[H]
\centering
\includegraphics[scale=0.83]{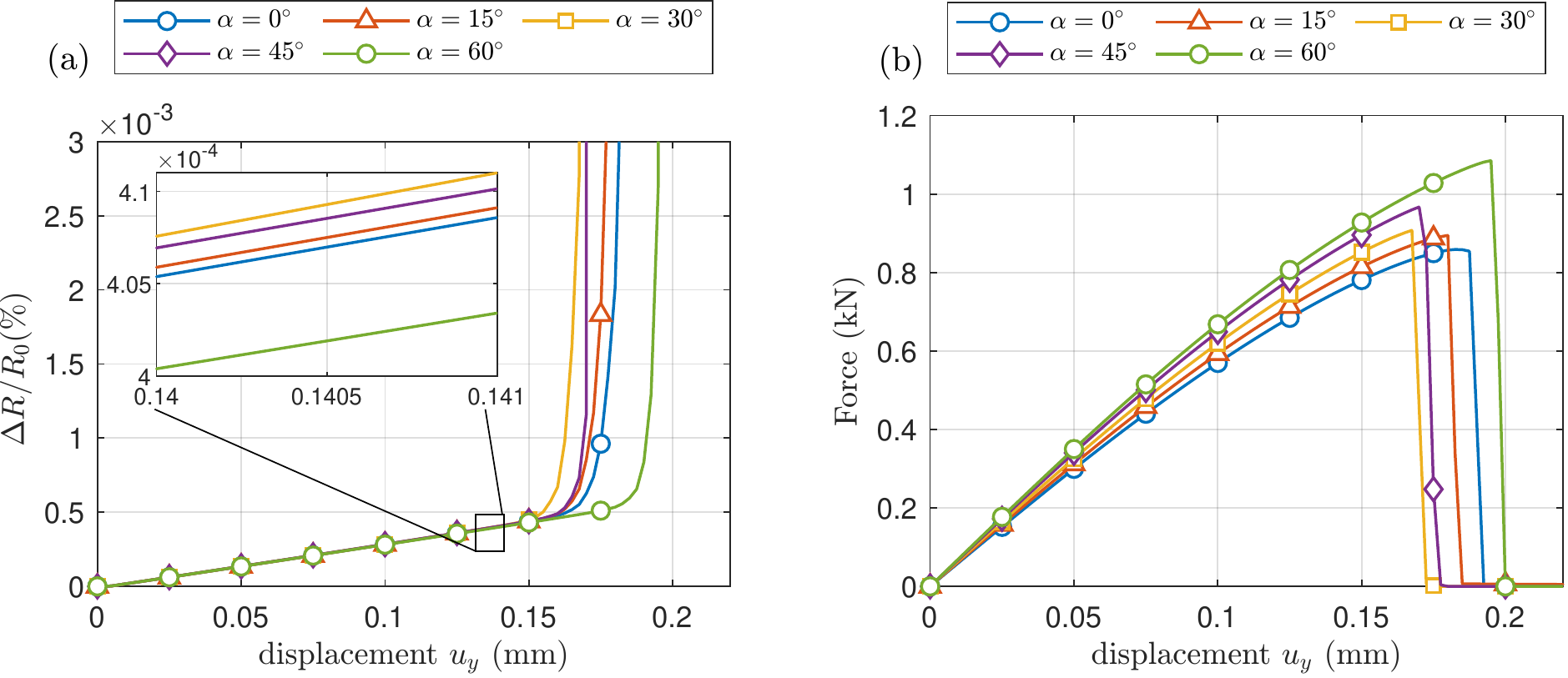}
\caption{Effect of the crack inclination angle on: (a) the relative resistance $\Delta R/R_{0}$ and (b) load-displacement curves.} 
\label{Fig:case_2_angle}
\end{figure}

The effect of the crack inclination angle is investigated in terms of the relative variation of the electrical resistance and the load-displacement curve of the plate in Figs.~\ref{Fig:case_2_angle}a and b, respectively. The results show that there are some critical angles that induce an early breakage of the plate as a result of the combination of the stress concentrations around the notch and the holes. This is the case of $\alpha=30^{\circ{}}$, which leads to a premature interruption of the current flow throughout the material as shown in Fig.~\ref{Fig:case_2_angle}a. In this case, the crack propagates at an imposed displacement of $u_{y}=0.1725$ mm, growing across the plate along the direction of maximum energy release rate, as shown in Fig. \ref{Fig:case_2}.

\subsection{Fracture of a thin plate containing a random distribution of defects}
\label{Section4:case3}

\begin{figure}[H]
\centering
\includegraphics[scale=0.9]{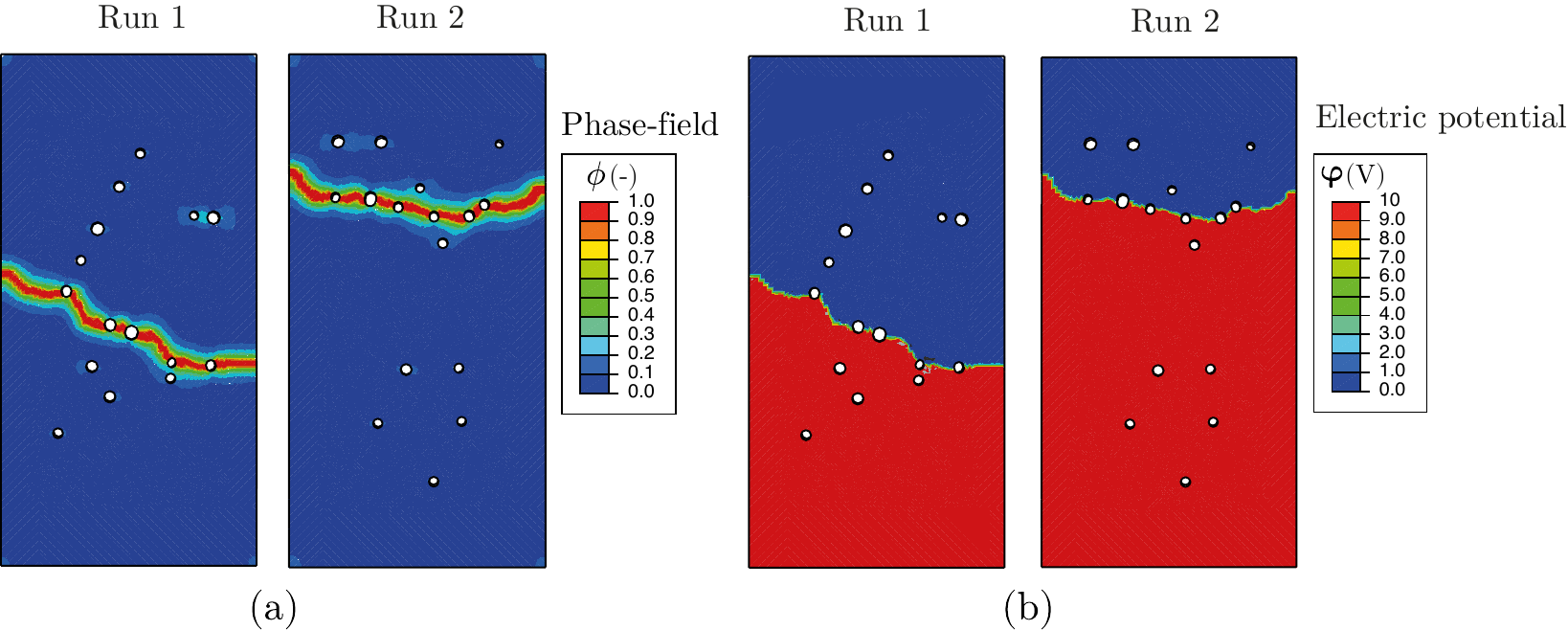}
\caption{Contour plots after crack propagation of two representative simulations of CNT/epoxy composite plates with a random distributions of defects: (a) phase-field $\phi$, and (b) electric potential $\varphi$.}
\label{Fig:case_3}
\end{figure}

\begin{figure}[H]
\centering
\includegraphics[scale=0.78]{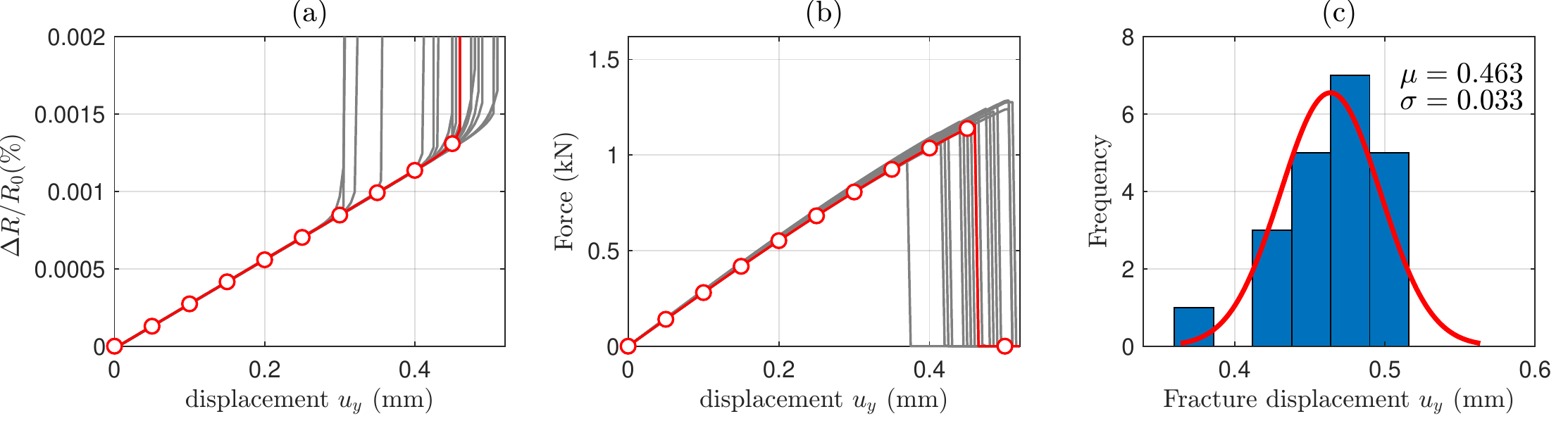}
\caption{Fracture of a plate containing a random distribution of defects: (a) probabilistic analysis of the relative variation of electrical resistance resistance $\Delta R/R_{0}$, (b) load-displacement curves of CNT/epoxy composite plates with random distribution of defects, and (c) the corresponding histogram of the ultimate fracture displacement. Grey and red lines in (a) and (b) correspond to the conducted Monte Carlo simulations and their mean values, respectively.} 
\label{Fig:case_3_plot}
\end{figure}

This case study investigates the electromechanical response of CNT-reinforced plates with a random distribution of defects, as sketched in Fig.~\ref{Fig:model}c. The holes are located in the area highlighted in red in Fig.~\ref{Fig:model}c, following a random uniform distribution until subtracting 1$\%$ of the total volume of the plate. The defects radii follows a normal distribution with mean and standard deviation values of 2 \rm{mm} and 1.2 \rm{mm}, respectively. The finite element model uses a total of 500,000 DOFs, with the phase-field length scale ($\ell = 0.0005$ mm) being five times larger than the characteristic element length. Figures~\ref{Fig:case_3}a and b show the contour plots of the phase-field and the electrical potential after the failure of two sample plates with different distributions of defects. The uncertainty in the electromechanical response is quantified through direct Monte Carlo simulations in terms of relative variation of resistance-displacement and force-displacement curves, as shown in Figs.~\ref{Fig:case_3_plot}a and b. A total of 21 simulations have been conducted, and the histogram of the ultimate fracture displacements is reported in Figure~\ref{Fig:case_3_plot}c. These results demonstrate the flexibility of the proposed approach to simulate crack initiation and propagation of piezoresistive materials with arbitrary crack patterns, allowing to conduct uncertainty propagation analyses without time-consuming mesh adaptation requirements.

\subsection{3D crack growth in a cracked cylinder}
\label{Section4:case5}

\begin{figure}[H]
\centering
\includegraphics[scale=1.00]{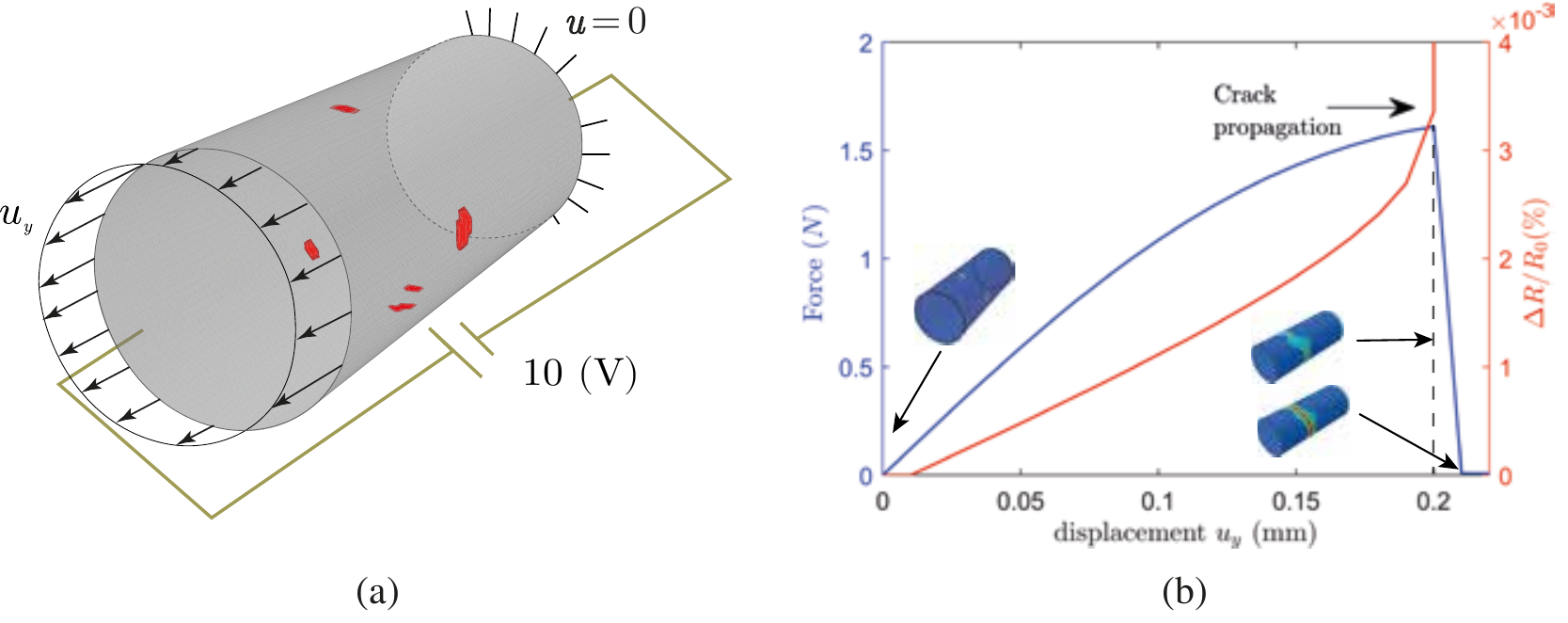}
\caption{Boundary conditions of the 3D case study of a CNT/epoxy cylinder: (a) initial crack distribution, as highlighted with red colour, and (b) three-axis plot reporting the relative variation of electrical resistance and load-displacement curves.} 
\label{Fig:3D_1}
\end{figure}

\begin{figure}[H]
\centering
\includegraphics[scale=0.85]{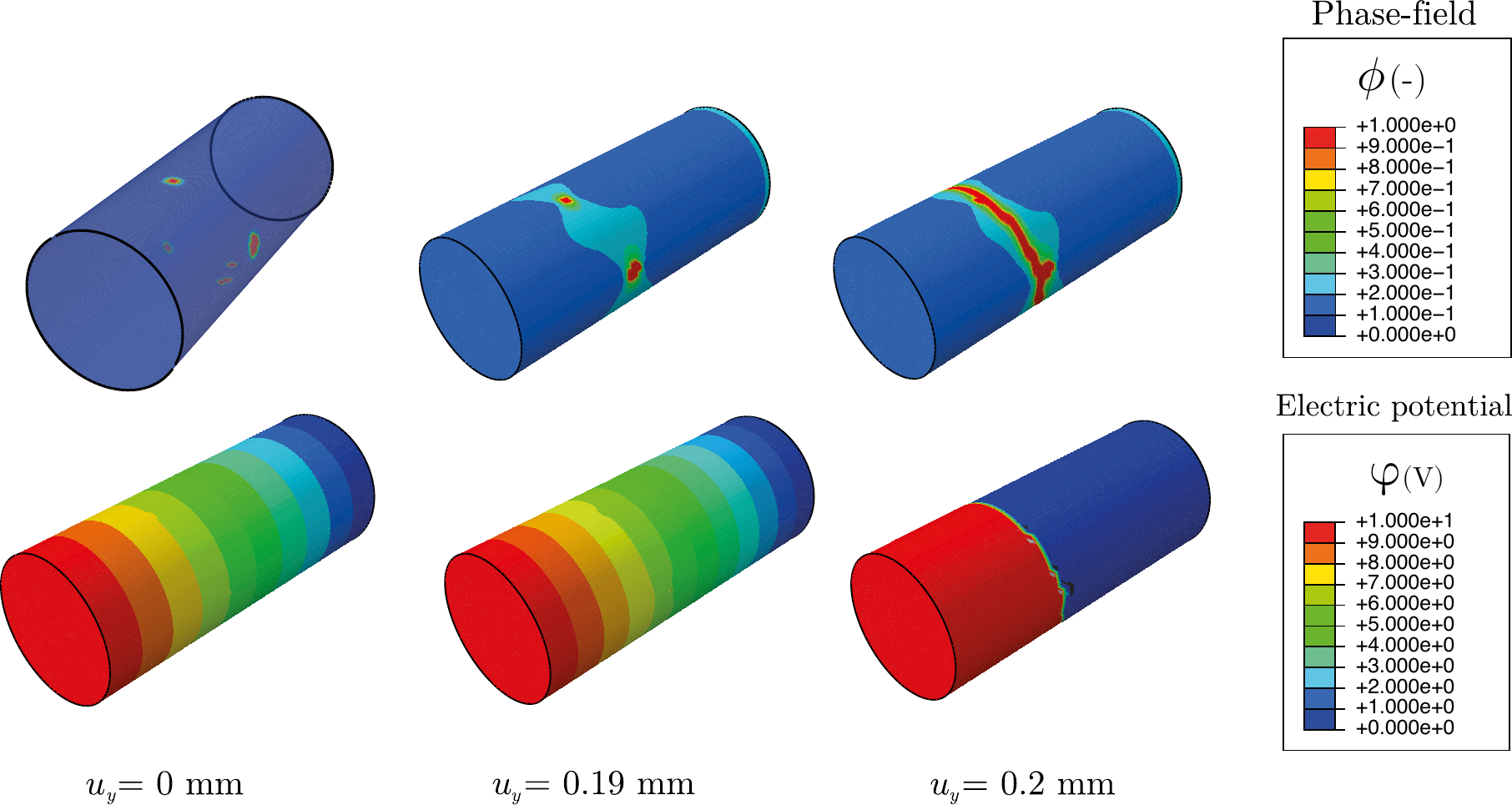}
\caption{3D crack growth in a cracked cylinder. Contours of crack evolution, as denoted by the phase-field $\phi$, and electric potential $\varphi$ in a 3D CNT/epoxy composite cylinder.} 
\label{Fig:3D_2}
\end{figure}

This last case study is aimed at illustrating the ability of the proposed approach for simulating the electromechanical response of self-sensing piezoresistive materials with arbitrary geometries and complex crack propagation patterns. Specifically, a three-dimensional cylinder with a radius of 2 $\rm{cm}$ and a length of 5 $\rm{cm}$ is investigated, as illustrated in Fig.~\ref{Fig:3D_1}a. The sample is pinned at one end, while a controlled displacement is imposed at the other end. A potential difference of 10 V is also imposed between the two bases of the cylinder. Five random notches are placed on the surface of the cylinder by defining the phase-field variable equal to $\phi=1$ as an initial condition (see Fig.~\ref{Fig:3D_2}a). The material properties used for this case study are those from Table~\ref{fittedparam} with a CNT volume fraction of 1\%. The finite element mesh comprises approximately 320,000 DOFs, with the characteristic element length being at least four time smaller than the phase-field length scale ($\ell = 0.001$ mm). The evolution of the phase-field variable and the electric potential are shown in Fig.~\ref{Fig:3D_2}. It can be readily observed how the phase-field variable starts propagating around the areas of the notches to coalesce and finally cross the entire cross-section of the structure. The corresponding force-displacement curve and the relative variation of the electrical resistance of the structure are depicted in the 3-axis plot of Fig.~\ref{Fig:3D_1}b. It is shown in this figure that the relative variation of the electrical resistance first exhibits a quasi-linear behaviour dominated by piezoresistance before the defects start to propagate. Once the defects start propagating, the electrical resistance starts to raise in a non-linear way, according to the implemented degradation function. Finally, once the crack crosses completely the structure, the current flow between the electrodes is interrupted and the electrical resistance tends to infinite.

\section{Concluding remarks}
\label{Section:5-Conclusion}

We have presented a novel formulation to simulate electromechanical fracture in piezoresistive composite materials. The model combines mean field homogenization (MFH) and phase-field fracture within a structural-electrical framework. The formulation has been particularized to CNT-based composites, and a complete micromechanics framework has been used to estimate the effective constitutive properties of the composite material, including the elastic tensor, electrical conductivity, and linear piezoresistivity tensor. The proposed approach allows relating the macroscopic response to the fundamental features of the composite microstructure such as the filler volume fraction and geometry, or to the individual material properties of the constituent phases. Then, the governing equations of fracture of linear piezoresistive materials using linear electromechanics and the phase-field method have been derived. The proposed formulation has been numerically implemented using the finite element method, and the resulting code is made freely available to the scientific community. To assess the accuracy and capabilities of the proposed approach, detailed parametric analyses and five different case studies of increasing complexity have been presented. The presented numerical results have demonstrated the accuracy and flexibility of the proposed approach to predict the electromechanical response of smart piezoresistive structures with general geometries and experiencing arbitrary crack propagation patterns. The presented formulation is envisaged to serve as a valuable computational tool to generate accurate digital twins with large applicability for design optimisation of self-diagnostic composites and signal processing for SHM applications.

\section*{Acknowledgements}
\label{Sec:Acknowledgeoffunding}

L. Quinteros acknowledges financial support from the National Agency for Research and Development (ANID)/  Scholarship Program / DOCTORADO BECAS CHILE/2020 - 72210161. E. Garc\'{\i}a-Mac\'{\i}as was supported by the Consejería de Transformación Económica, Conocimiento, Empresas y Universidades de la Junta de Andalucía (Spain) through the research project P18-RT-3128. E. Mart\'{\i}nez-Pa\~neda was supported by an UKRI Future Leaders Fellowship (grant MR/V024124/1).

\appendix
\section{CNT-based composite model}
\label{Apendix_A}
Two mechanisms govern the electrical conductivity in CNT-based composites, namely electro hopping (EH) and conductive networking (CN). The probability of electro hopping depends on the average distance between tubes $d_{a,\chi} (\bm{\varepsilon})$, which has been reported to follow a power-low relationship as~\cite{FENG2013143},
\begin{equation}
d_{a, \chi}(\varepsilon)= \begin{cases}d_{c} & \chi=E H \\ d_{c}\left(\frac{f_{c}(\mathbf{\varepsilon)}}{f(\mathbf{\varepsilon)}}\right)^{1 / 3} & \chi=C N\end{cases}
\end{equation}
\noindent with $d_{c}$ being the maximum separation between CNTs that allows the electron transfer. This effect can be modelled with a continuum interphase layer coating the CNTs, using the generalized Simmons formula as~\cite{Simmons1963b},
\begin{equation}
R_{i n t, \chi}\left(\bm{\varepsilon}, d_{a, \chi}(\bm{\varepsilon})\right)=\frac{d_{a, \chi}(\bm{\varepsilon}) \hbar^{2}}{a e^{2}(2 m \lambda)^{1 / 2}} \exp \left[\frac{4 \pi d_{a, \chi}(\bm{\varepsilon})}{\hbar}(2 m \lambda)^{1 / 2}\right]
\end{equation}

\noindent where $m$ and $e$ are the mass and electric charge of an electron, $\lambda$ is the height of the tunneling potential barrier, $a$ is the contact area of the CNTs, and $\hbar$ is the reduced Plank's constant. The thickness of the conductive interphase and its electrical conductivity is given by~\cite{Seidel2009},
 \begin{equation}
t_{\chi}=\frac{1}{2} d_{a, \chi}(\bm{\varepsilon}), \quad \sigma_{i n t, \chi}=\frac{d_{a, \chi}(\bm{\varepsilon})}{a R_{i n t, \chi}\left(\bm{\varepsilon}, d_{a, \chi}(\bm{\varepsilon})\right)}.
\end{equation}

The proposed interphase layer is modelled as an effective composite solid cylinder. Therefore the effective conductivity tensor $\sigma_{eff}$ is defined as transversely isotropic with effective longitudinal and transverse electrical conductivities, denoted by $\tilde{\sigma}_{\chi}^{L}$ and $\tilde{\sigma}_{\chi}^{T}$, respectively. Then, applying Maxwell's equations and the rule of mixtures,
\begin{equation}
\tilde{\sigma}_{\chi}^{L}(\bm{\varepsilon})=\frac{\left(L+2 t_{\chi}(\bm{\varepsilon})\right) \sigma_{i n t, \chi}(\bm{\varepsilon})\left[\sigma_{c}^{L} r_{c}^{2}+\sigma_{i n t, \chi}(\bm{\varepsilon})\left(2 r_{c} t_{\chi}(\bm{\varepsilon})+t_{\chi}^{2}(\bm{\varepsilon})\right)\right]}{2 \sigma_{c}^{L} r_{c}^{2} t_{\chi}(\bm{\varepsilon})+2 \sigma_{i n t, \chi}(\bm{\varepsilon})\left(2 r_{c} t_{\chi}(\bm{\varepsilon})+t_{\chi}^{2}(\bm{\varepsilon})\right) t_{\chi}(\bm{\varepsilon})+\sigma_{i n t, \chi}(\bm{\varepsilon}) L\left(r_{c}+t_{\chi}(\bm{\varepsilon})\right)^{2}} 
\end{equation}
\begin{equation}
\tilde{\sigma}_{\chi}^{T}(\bm{\varepsilon})=\frac{\sigma_{i n t, \chi}(\bm{\varepsilon})}{L+2 t_{\chi}(\bm{\varepsilon})}\left[L \frac{2 r_{c}^{2} \sigma_{c}^{T}+\left(\sigma_{c}^{T}+\sigma_{i n t, \chi}(\bm{\varepsilon})\right)\left(t_{\chi}^{2}(\bm{\varepsilon})+2 r_{c} t_{\chi}(\bm{\varepsilon})\right)}{2 r_{c}^{2} \sigma_{i n t, \chi}(\bm{\varepsilon})+\left(\sigma_{c}^{T}+\sigma_{i n t, \chi}(\bm{\varepsilon})\right)\left(t_{\chi}^{2}(\bm{\varepsilon})+2 r_{c} t_{\chi}(\bm{\varepsilon})\right)}+2 t_{\chi}(\bm{\varepsilon})\right].
\end{equation}
The resultant filler is larger than the original, due to the interphase, and thus it must be updated as, 
\begin{equation}
f_{e f f, \chi}(\bm{\varepsilon})=\frac{\left(r_{c}+t_{\chi}(\bm{\varepsilon})\right)^{2}\left(L+2 t_{\chi}(\bm{\varepsilon})\right)}{r_{c}^{2} L} f(\bm{\varepsilon}).
\end{equation}

%% BIBLIOGRAPHY

\bibliographystyle{elsarticle-num-nodoi}

%{\footnotesize
%\bibliography{bibliographyCNT,library}}

\end{document}